\documentclass[english,letterpaper,prd,showpacs,preprintnumbers,nofootinbib,floatfix,preprint,tightenlines]{revtex4}
\usepackage[T1]{fontenc}
\usepackage[latin1]{inputenc}
\usepackage{graphicx}

\makeatletter

\makeatletter

\usepackage{latexsym}

\usepackage{epsfig}

\makeatother

\usepackage{babel}
\makeatother
\begin{document}
\newcommand{\barl}{\bar{\lambda}}

\newcommand{\barp}{\bar{p}}

\preprint{ANL-HEP-PR-07-5, hep-ph/0702003}

\title{Gluon-gluon contributions to the production of continuum diphoton
pairs at hadron colliders}

\author{P. M. Nadolsky,$^{1}$%
\thanks{nadolsky@hep.anl.gov%
}~C. Bal\'{a}zs,$^{1}$%
\thanks{balazs@hep.anl.gov%
} E. L. Berger,$^{1}$%
\thanks{berger@hep.anl.gov%
}~ and C.-P. Yuan$^{2}$%
\thanks{yuan@pa.msu.edu%
}}

\affiliation{$^{1}$High Energy Physics Division, Argonne National Laboratory,
Argonne, IL 60439, USA \\
 $^{2}$Department of Physics and Astronomy, Michigan State University,
East Lansing, MI 48824, USA}

\begin{abstract}
We compute the contributions to continuum photon pair production at
hadron colliders from processes initiated by gluon-gluon and gluon-quark
scattering into two photons through a four-leg virtual quark loop.
Complete two-loop cross sections in perturbative quantum chromodynamics
are combined with contributions from soft parton radiation resummed
to all orders in the strong coupling strength. The structure of the
resummed cross section is examined in detail, including a new type
of unintegrated parton distribution function affecting azimuthal angle
distributions of photons in the pair's rest frame. As a result of
this analysis, we predict diphoton transverse momentum distributions
in gluon-gluon scattering in wide ranges of kinematic parameters at
the Fermilab Tevatron and the CERN Large Hadron Collider. 
\end{abstract}

\date{January 31, 2007}



\maketitle

\section{Introduction}

Advances in the computation of higher-order radiative contributions
in perturbative quantum chromodynamics (PQCD) open opportunities to
predict hadronic observables at an unprecedented level of precision.
Full realization of this potential requires concurrent improvements
in the methods for QCD factorization and resummation of logarithmic
enhancements in hadronic cross sections in infrared kinematic regions.
All-orders resummation of logarithmic corrections, such as the resummation
of transverse momentum ($Q_{T}$) logarithms in Drell-Yan-like processes~\cite{Collins:1984kg},
is increasingly challenging in multi-loop calculations as a result
of algebraic complexity and new types of logarithmic singularities
associated with multi-particle matrix elements.

In this paper, we address new theoretical issues in $Q_{T}$ resummation
at two-loop accuracy. We focus on photon pair production, particularly
on the gluon-gluon subprocess, $gg\rightarrow\gamma\gamma$, one of
the important short-distance subprocesses that contribute to the inclusive
reactions $p\bar{p}\rightarrow\gamma\gamma X$ at the Fermilab Tevatron
and $pp\rightarrow\gamma\gamma X$ at the CERN Large Hadron Collider
(LHC). This hadronic reaction is interesting in its own right, and
it is relevant in searches for the Higgs boson $h$, where it constitutes
an important QCD background to the $pp\rightarrow hX\rightarrow\gamma\gamma X$
production chain \cite{ATLAS:1999fr,CMSTDR:2006,Abdullin:2005yn}.
A reliable prediction of the cross section for $gg\rightarrow\gamma\gamma$
is needed for complete estimates of the $\gamma\gamma$ production
cross sections, a task that we pursue in accompanying papers~\cite{Balazs:2006cc,Balazs:2007hr}.

\begin{figure}
\includegraphics[width=12cm,keepaspectratio]{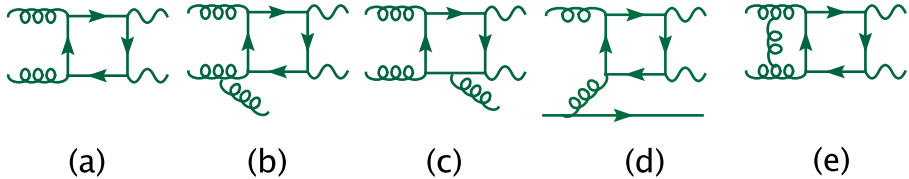}

\caption{Representative parton scattering subprocesses for diphoton production
in gluon-gluon scattering. \label{Fig:FeynDiag}}
\end{figure}

The lowest-order contribution to the cross section for $gg\rightarrow\gamma\gamma$
arises from a $2\rightarrow2$ diagram of order ${\mathcal{O}}(\alpha^{2}\alpha_{s}^{2})$
involving a 4-vertex virtual quark loop {[}Fig.~\ref{Fig:FeynDiag}(a)].
We evaluate all next-to-leading (NLO) contributions of order ${\mathcal{O}}(\alpha^{2}\alpha_{s}^{3})$
to the $gg\rightarrow\gamma\gamma$ process shown in Figs.~\ref{Fig:FeynDiag}(b-e).
An important new ingredient in this paper is the inclusion of the
$gq\rightarrow\gamma\gamma q$ process, Fig.~\ref{Fig:FeynDiag}(d),
a necessary component of the resummed NLO contribution. Our complete
treatment of the NLO cross section represents an improvement over
our original publication~\cite{Balazs:1997hv}, in which the large-$Q_{T}$
behavior of the $gg$ subprocess was approximated, and the $gq$ contribution
was not included. Furthermore, we resum to next-to-next-to-leading
logarithmic (NNLL) accuracy the large logarithmic terms of the form
$\ln(Q_{T}^{2}/Q^{2})$ in the limit when $Q_{T}$ of the $\gamma\gamma$
pair is much smaller than its invariant mass $Q$. Our NNLL cross
section includes the exact ${\mathcal{C}}$ coefficients of order
$\alpha_{s}$ for $gg+gq\rightarrow\gamma\gamma X$, and the functions
${\mathcal{A}}$ and ${\mathcal{B}}$ of orders $\alpha_{s}^{3}$
and $\alpha_{s}^{2}$ in all subprocesses, with these functions defined
in Sec.~\ref{Sec:Theory}.

We begin in Sec.~II with a summary of kinematics and our notation,
and we outline the partonic subprocesses that contribute to $\gamma\gamma$
production. In this section, we also derive a matrix element for the
$qg\rightarrow\gamma\gamma g$ process shown in Fig.~\ref{Fig:FeynDiag}(d),
a subprocess whose contribution is required to obtain consistent resummed
predictions for all values of $Q_{T}$. We obtain the ${\mathcal{O}}(\alpha^{2}\alpha_{s}^{3})$
cross section for the $gq\rightarrow\gamma\gamma q$ process from
the color-decomposed $q\bar{q}ggg$ amplitudes in Ref.~\cite{Bern:1994fz}.

The rich helicity structure of the $gg\rightarrow\gamma\gamma$ matrix
element is addressed in Sec.~III. The helicity dependence requires
a new type of transverse-momentum dependent (TMD) parton distribution
function (PDF) associated with the interference of amplitudes for
initial-state gluons of opposite helicities. The existence of the
helicity-flip TMD PDF modifies the azimuthal angle distributions of
the final-state photons, an effect that could potentially be observed
experimentally. By contrast, in vector boson production $pp\!\!\!{}^{{}^{(-)}}\rightarrow VX$
(with $V=\gamma^{*},W,Z,...$), such helicity-flip contributions are
suppressed as a result of the simple spin structure of the lowest-order
$q\bar{q}V$ coupling. In this section, we establish the presence
of helicity interference in the finite-order $2\rightarrow3$ cross
sections by systematically deriving their soft and collinear limits
in the splitting amplitude formalism \cite{Bern:1993mq,Bern:1993qk,Bern:1994zx,Bern:1994fz,Bern:1998sc,Bern:1999ry,Kosower:1999rx}.
We show how the helicity-flip TMD PDF arises from the general structure
of the small-$Q_{T}$ resummed cross section.

Section~IV contains some numerical predictions for the Tevatron and
LHC, where we show the fraction of the rate for $\gamma\gamma$ production
supplied by the $gg+gq$ subprocess. The generally expected prominence
of $gg+gq$ scattering at the LHC is only partially supported by our
findings. The large $gg$ partonic luminosity cannot fully compensate
for the small cross section associated with $gg$ scattering. Our
findings are summarized in Sec.~V. Three Appendices are included.
In Appendix~\ref{Appendix:qgAAq}, we present some of the details
of our derivation of the amplitude for the subprocess $qg\rightarrow\gamma\gamma g$.
In Appendix~\ref{Appendix:ASYgg}, we derive the small-$Q_{T}$ asymptotic
form of the NLO cross section for $gg\rightarrow\gamma\gamma$.

\section{Notation and Subprocesses \label{Sec:Theory}}

\subsection{Notation \label{sub:Notations}}

We consider the scattering process $h_{1}(P_{1})+h_{2}(P_{2})\rightarrow\gamma(P_{3})+\gamma(P_{4})+X$,
where $h_{1}$ and $h_{2}$ are the initial-state hadrons. In terms
of the center-of-mass collision energy $\sqrt{S}$, the $\gamma\gamma$
invariant mass $Q$, the $\gamma\gamma$ transverse momentum $Q_{T}$,
and the $\gamma\gamma$ rapidity $y$, the momenta $P_{1}^{\mu}$
and $P_{2}^{\mu}$ of the initial hadrons and $q^{\mu}\equiv P_{3}^{\mu}+P_{4}^{\mu}$
of the pair are expressed in the laboratory frame as\begin{eqnarray}
P_{1}^{\mu} & = & \frac{\sqrt{S}}{2}\left\{ 1,0,0,1\right\} ;\\
P_{2}^{\mu} & = & \frac{\sqrt{S}}{2}\left\{ 1,0,0,-1\right\} ;\\
q^{\mu} & = & \left\{ \sqrt{Q^{2}+Q_{T}^{2}}\cosh y,Q_{T},0,\sqrt{Q^{2}+Q_{T}^{2}}\sinh y\right\} .\end{eqnarray}
 The light-cone momentum fractions for the boosted $2\rightarrow2$
scattering system are\begin{equation}
x_{1,2}\equiv\frac{2(P_{2,1}\cdot q)}{S}=\frac{\sqrt{Q^{2}+Q_{T}^{2}}e^{\pm y}}{\sqrt{S}}.\end{equation}

Decay of the $\gamma\gamma$ pairs is described in the hadronic Collins-Soper
frame \cite{Collins:1977iv}. The Collins-Soper frame is a rest frame
of the $\gamma\gamma$ pair (with $q^{\mu}=\left\{ Q,0,0,0\right\} $
in this frame), chosen so that (a) the momenta $\vec{P}_{1}$ and
$\vec{P}_{2}$ of the initial hadrons lie in the $Oxz$ plane (with
zero azimuthal angle), and (b) the $z$ axis bisects the angle between
$\vec{P}_{1}$ and $-\vec{P}_{2}$. The photon momenta are antiparallel
in the Collins-Soper frame: \begin{eqnarray}
P_{3}^{\mu} & = & \frac{Q}{2}\left\{ 0,\sin\theta_{*}\cos\varphi_{*},\sin\theta_{*}\sin\varphi_{*},\cos\theta_{*}\right\} ,\label{p3CS}\\
P_{4}^{\mu} & = & \frac{Q}{2}\left\{ 0,-\sin\theta_{*}\cos\varphi_{*},-\sin\theta_{*}\sin\varphi_{*},-\cos\theta_{*}\right\} ,\label{p4CS}\end{eqnarray}
 where $\theta_{*}$ and $\varphi_{*}$ are the photon's polar and
azimuthal angles. Our aim is to derive resummed predictions for the
fully differential $\gamma\gamma$ cross section $d\sigma/(dQ^{2}dydQ_{T}^{2}d\Omega_{*}),$
where $d\Omega_{*}=d\cos\theta_{*}d\varphi_{*}$ is a solid angle
element around the direction of $\vec{P}_{3}$ in the Collins-Soper
frame of reference defined in Eq.~(\ref{p3CS}). The parton momenta
and helicities are denoted by lowercase $p_{i}$ and $\lambda_{i}$.

\subsection{Scattering contributions \label{subsection:OverviewFeynmanDiagrams}}

We concentrate on direct production of isolated photons in hard QCD
scattering, the dominant production process at hadron colliders. A
number of hard-scattering contributions to the processes $q\bar{q}+qg\rightarrow\gamma\gamma,$
as well as photon production via fragmentation, have been studied
in the past \cite{Aurenche:1985yk,Bailey:1992br,Binoth:1999qq}. Our
numerical calculations include the lowest-order process $q\bar{q}\rightarrow\gamma\gamma$
of order ${\cal O}(\alpha^{2})$ and contributions from $q\bar{q}\rightarrow\gamma\gamma g$
and $q^{\!\!\!\!\!{}^{(-)}}g\rightarrow\gamma\gamma q^{\!\!\!\!\!{}^{(-)}}$
of order ${\cal O}(\alpha^{2}\alpha_{s}),$ where $\alpha(\mu)=e^{2}/4\pi$
and $\alpha_{s}(\mu)=g^{2}/4\pi$ are the running QED and QCD coupling
strengths.

Glue-glue scattering is the next leading direct production channel,
with the full set of NLO contributions shown in Fig.~\ref{Fig:FeynDiag}.
Production of $\gamma\gamma$ pairs via a box diagram in $gg$ scattering
as in Fig.~\ref{Fig:FeynDiag}(a) \cite{Berger:1983yi} is suppressed
by two powers of $\alpha_{s}$ compared to the lowest-order $q\bar{q}\rightarrow\gamma\gamma$
contribution, but is enhanced by a product of two large gluon PDF's
if typical momentum fractions $x$ are small. The main ${\mathcal{O}}(\alpha^{2}\alpha_{s}^{3})$,
or NLO, corrections, include one-loop $gg\rightarrow\gamma\gamma g$
diagrams (b) and (c) derived in \cite{Balazs:1999yf,deFlorian:1999tp},
as well as 4-leg two-loop diagrams (e) computed in \cite{Bern:2001df}.
The real and virtual diagrams are combined in Ref.~\cite{Bern:2002jx}
to obtain the full NLO contribution from $gg$ scattering. In this
study we also include subleading NLO contributions from the process
(d), $gq_{S}\rightarrow\gamma\gamma q_{S}$ via the quark loop, where
$q_{S}=\sum_{i=u,d,s,...}(q_{i}+\bar{q}_{i})$ denotes the flavor-singlet
combination of quark scattering channels. The $gq_{S}\rightarrow\gamma\gamma q_{S}$
helicity amplitude is derived from the one-loop $q\bar{q}ggg$ amplitude
\cite{Bern:1994fz} and explicitly presented in Appendix~\ref{Appendix:qgAAq}.
As a cross check, we verified that this amplitude correctly reproduces
the known collinear limits. Our result does not confirm an expression
for this amplitude available in the literature \cite{Yasui:2002bn},
which does not satisfy these limits. When evaluated in our resummation
calculation under typical event selection conditions, $gg+gq_{S}$
scattering contributes about 20\% and 10\% of the total rate at the
LHC and the Tevatron, respectively, but this fraction can be larger
in specific regions of phase space.

\section{Theoretical Presentation}

\subsection{Small-$Q_{T}$ asymptotics of the next-to-leading order cross section\label{subsection:AsymptoticISR}}

When the transverse momentum $Q_{T}$ of the diphoton approaches zero,
the NLO production cross section $d\sigma/(dQ^{2}dy\, dQ_{T}^{2}d\Omega_{*})$,
or briefly $P(Q,Q_{T},y,\Omega_{*}),$ is dominated by $\gamma\gamma$
recoil against soft and collinear QCD radiation. In this subsection
we concentrate on the effects of initial-state QCD radiation and derive
the leading small-$Q_{T}$ part of the NLO differential cross section,
called the asymptotic term $A(Q,Q_{T},y,\Omega_{*})$.

The ${\mathcal{O}}(\alpha_{s})$ asymptotic cross section valid at
$Q_{T}^{2}\ll Q^{2}$ consists of a few generalized functions that
are integrable on an interval $0\leq Q_{T}\leq P_{T}$, with $P_{T}$
being a finite value of transverse momentum: \begin{eqnarray}
 &  & A(Q,Q_{T},y,\Omega_{*})=F_{\delta}(Q,y,\Omega_{*})\delta(\vec{Q}_{T})\nonumber \\
 & + & F_{1}(Q,y,\Omega_{*})\left[\frac{1}{Q_{T}^{2}}\ln\frac{Q^{2}}{Q_{T}^{2}}\right]_{+}+F_{0}(Q,y,\Omega_{*})\left[\frac{1}{Q_{T}^{2}}\right]_{+}+\dots.\label{ASYgeneric}\end{eqnarray}
 The {}``$+$'' prescription $\left[f(Q_{T})\right]_{+}$ is defined
for a function $f(Q_{T})$ and a smooth function $g(Q_{T})$ as\begin{eqnarray}
\int_{0}^{P_{T}^{2}}dQ_{T}^{2}\left[f(Q_{T})\right]_{+}g(Q_{T}) & \equiv & \int_{0}^{P_{T}^{2}}dQ_{T}^{2}f(Q_{T})\left(g(Q_{T})-g(0)\right);\\
\left[f(Q_{T})\right]_{+}=f(Q_{T}) & \mbox{ for} & Q_{T}\neq0.\end{eqnarray}
 Subleading terms proportional to $\left(Q/Q_{T}\right)^{p}$ with
$p\leq1$ are neglected in Eq.~(\ref{ASYgeneric}). Its form is influenced
by spin correlations between the initial-state partons and final-state
photons. As a consequence of these spin correlations, the functions
$F_{\delta},$ $F_{0}$, and $F_{1}$ depend on the direction of the
final-state photons in the Collins-Soper frame (the polar angle $\theta_{*}$
and sometimes the azimuthal angle $\varphi_{*}$).

The spin dependence of the small-$Q_{T}$ cross section in the $gg\rightarrow\gamma\gamma g$
and $gq_{S}\rightarrow\gamma\gamma q_{S}$ channels is complex. The
Born-level process $g(p_{1},\lambda_{1})+g(p_{2},\lambda_{2})\rightarrow\gamma(p_{3},\lambda_{3})+\gamma(p_{4},\lambda_{4})$
is described by 16 non-zero helicity amplitudes ${\mathcal{M}}_{4}(p_{1},\lambda_{1};p_{2},\lambda_{2};p_{3},\lambda_{3};p_{4},\lambda_{4})\equiv{\mathcal{M}}_{4}(\lambda_{1},\lambda_{2},\lambda_{3},\lambda_{4})$
for quark-box diagrams of the type shown in Fig.~\ref{Fig:FeynDiag}(a).
The normalization of ${\mathcal{M}}_{4}(\lambda_{1},\lambda_{2},\lambda_{3},\lambda_{4})$
is chosen so that the unpolarized Born $gg\rightarrow\gamma\gamma$
cross section reads as\begin{equation}
\left.\frac{d\sigma_{gg}}{dQ^{2}dy\, dQ_{T}^{2}d\Omega_{*}}\right|_{Born}=\delta(\vec{Q}_{T})\frac{\Sigma_{g}(\theta_{*})}{S}f_{g/h_{1}}(x_{1},\mu_{F})f_{g/h_{2}}(x_{2},\mu_{F}),\label{Borngg}\end{equation}
 where\begin{equation}
\Sigma_{g}(\theta_{*})\equiv\sigma_{g}^{(0)}L_{g}(\theta_{*}),\label{Sigmag}\end{equation}
 with\begin{equation}
\sigma_{g}^{(0)}=\frac{\alpha^{2}(Q)\alpha_{s}^{2}(Q)}{32\pi Q^{2}(N_{c}^{2}-1)}\left(\sum_{i}e_{i}^{2}\right)^{2},\end{equation}
 and \begin{equation}
L_{g}(\theta_{*})\equiv\sum_{\lambda_{1},\lambda_{2},\lambda_{3},\lambda_{4}=\pm1}\left|{\mathcal{M}}_{4}(\lambda_{1},\lambda_{2},\lambda_{3},\lambda_{4})\right|^{2}.\label{L}\end{equation}
 In these equations, $N_{c}=3$ is the number of QCD colors, $e_{i}$
is the fractional electric charge (in units of the positron charge
$e$) of the quark $i$ circulating in the loop, and $f_{g/h}(x,\mu_{F})$
is the gluon PDF evaluated at a factorization scale $\mu_{F}$. The
right-hand side of Eq.~(\ref{L}) includes summation over gluon and
photon helicities $\lambda_{i}$, with $i=1,...,4$.

\begin{figure}
\begin{centering}
\includegraphics[width=1\textwidth,height=8cm,keepaspectratio]{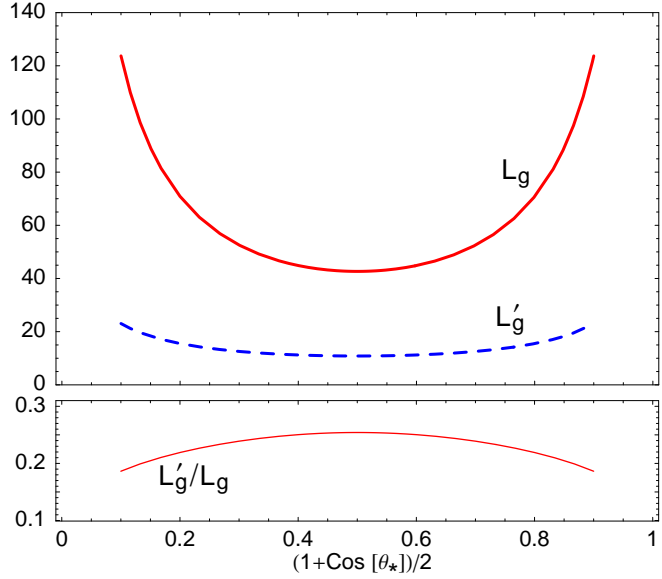}
\par\end{centering}

\caption{The functions $L_{g}(\theta_{*})$ and $L_{g}^{\prime}(\theta_{*})$
arising in the $gg\rightarrow\gamma\gamma$ asymptotic cross section
(\ref{ASYgg}) and their ratio. \label{Fig:LLprime}}
\end{figure}

At NLO, the small-$Q_{T}$ cross section is proportional to the angular
function $\Sigma_{g}(\theta_{*})$ (the same as in the Born cross
section), and another function \begin{eqnarray}
\Sigma_{g}^{\prime}(\theta_{*},\varphi_{*}) & = & \sigma_{g}^{(0)}\sum_{\lambda_{1},\lambda_{2},\lambda_{3},\lambda_{4}=\pm1}{\mathcal{M}}_{4}^{*}(\lambda_{1},\lambda_{2},\lambda_{3},\lambda_{4}){\mathcal{M}}_{4}(-\lambda_{1},\lambda_{2},\lambda_{3},\lambda_{4})\nonumber \\
 & \equiv & \sigma_{g}^{(0)}L_{g}^{\prime}(\theta_{*})\cos2\varphi_{*}.\label{Sigmagprime}\end{eqnarray}
 The function $\Sigma_{g}^{\prime}(\theta_{*},\varphi_{*})$ is obtained
by spin-averaging the product of the amplitude ${\mathcal{M}}_{4}(\lambda_{1},\lambda_{2},\lambda_{3},\lambda_{4}),$
and the complex-conjugate amplitude ${\mathcal{M}}_{4}^{*}(-\lambda_{1},\lambda_{2},\lambda_{3},\lambda_{4})$
evaluated with the reverse sign of the helicity $\lambda_{1}$. The
sign flip for $\lambda_{1}$ results in dependence of $\Sigma_{g}^{\prime}(\theta_{*},\varphi_{*})$
on $\cos2\varphi_{*}$. The $\theta_{*}$ dependence of $\Sigma_{g}^{\prime}(\theta_{*},\varphi_{*})$
enters through the function\begin{eqnarray}
L_{g}^{\prime}(\theta_{*}) & = & -4\mbox{{Re}}\left(M_{1,1,-1,-1}^{(1)}+M_{1,-1,1,-1}^{(1)}+M_{-1,1,1,-1}^{(1)}+1\right),\label{Lprime}\end{eqnarray}
 presented in terms of reduced amplitudes $M_{\lambda_{1},\lambda_{2},\lambda_{3},\lambda_{4}}^{(1)}$
in the notation of Ref.~\cite{Bern:2001df}. For comparison, the
functions $L_{g}(\theta_{*})$ and $L_{g}^{\prime}(\theta_{*})$ are
plotted versus $(1+\cos\theta_{*})/2$ in Fig.~\ref{Fig:LLprime}.

The NLO asymptotic term in the sum of the contributions from the $gg\rightarrow\gamma\gamma$
and $gq_{S}\rightarrow\gamma\gamma$ channels (denoted as $gg+gq_{S}$
channel) is \begin{eqnarray}
A(Q,Q_{T},y,\Omega_{*}) & = & \frac{1}{S}\Biggl\{\Sigma_{g}(\theta_{*})\left[\delta(\vec{Q}_{T})F_{g,\delta}(Q,y,\theta_{*})+F_{g,+}(Q,y,Q_{T})\right]\nonumber \\
 &  & \hspace{12pt}+\Sigma_{g}^{\prime}(\theta_{*},\varphi_{*})F_{g,+}^{\prime}(Q,y,Q_{T})\Biggr\}.\label{ASYgg}\end{eqnarray}
 Here\begin{eqnarray}
 &  & F_{g,\delta}\equiv f_{g/h_{1}}(x_{1},\mu_{F})f_{g/h_{2}}(x_{2},\mu_{F})\left(1+2\frac{\alpha_{s}}{\pi}h_{g}^{(1)}(\theta_{*})\right)\nonumber \\
 & + & \frac{\alpha_{s}}{\pi}\Biggl\{\left(\left[{\mathcal{C}}_{g/a}^{(1,c)}\otimes f_{a/h_{1}}\right](x_{1},\mu_{F})-\left[P_{g/a}\otimes f_{a/h_{1}}\right](x_{1},\mu_{F})\,\ln\frac{\mu_{F}}{Q}\right)\, f_{g/h_{2}}(x_{2},\mu_{F})\nonumber \\
 &  & \hspace{27pt}+f_{g/h_{1}}(x_{1},\mu_{F})\left(\left[{\mathcal{C}}_{g/a}^{(1,c)}\otimes f_{a/h_{2}}\right](x_{2},\mu_{F})-\left[P_{g/a}\otimes f_{a/h_{2}}\right](x_{2},\mu_{F})\,\ln\frac{\mu_{F}}{Q}\right)\Biggr\};\label{FgDelta}\end{eqnarray}
 \begin{eqnarray}
F_{g,+} & = & \frac{1}{2\pi}\frac{\alpha_{s}}{\pi}\Biggl\{ f_{g/h_{1}}(x_{1},\mu_{F})f_{g/h_{2}}(x_{2},\mu_{F})\left({\mathcal{A}}_{g}^{(1,c)}\left[\frac{1}{Q_{T}^{2}}\ln\frac{Q^{2}}{Q_{T}^{2}}\right]_{+}+{\mathcal{{\mathcal{B}}}}_{g}^{(1,c)}\left[\frac{1}{Q_{T}^{2}}\right]_{+}\right)\nonumber \\
 & + & \left[\frac{1}{Q_{T}^{2}}\right]_{+}\Bigl(\left[P_{g/a}\otimes f_{a/h_{1}}\right](x_{1},\mu_{F})\, f_{g/h_{2}}(x_{2},\mu_{F})\nonumber \\
 &  & \hspace{47pt}+f_{g/h_{1}}(x_{1},\mu_{F})\left[P_{g/a}\otimes f_{a/h_{2}}\right](x_{2},\mu_{F})\Bigr)\Biggr\};\label{FgPlus}\end{eqnarray}
 and\begin{eqnarray}
F_{g,+}^{\prime} & = & \frac{1}{2\pi}\frac{\alpha_{s}}{\pi}\left[\frac{1}{Q_{T}^{2}}\right]_{+}\Bigl(\left[P_{g/g}^{\prime}\otimes f_{g/h_{1}}\right](x_{1},\mu_{F})\, f_{g/h_{2}}(x_{2},\mu_{F})\nonumber \\
 &  & \hspace{77pt}+f_{g/h_{1}}(x_{1},\mu_{F})\left[P_{g/g}^{\prime}\otimes f_{g/h_{2}}\right](x_{2},\mu_{F})\Bigr).\label{FgPlusPrime}\end{eqnarray}
 The ${\cal O}(\alpha_{s}/\pi)$ coefficients ${\mathcal{A}}_{g}^{(1,c)}$,
${\mathcal{B}}_{g}^{(1,c)}$ and functions ${\mathcal{C}}_{g/a}^{(1,c)}(x,b)$,
$h_{g}^{(1)}(\theta_{*})$ are defined and listed explicitly in Ref.~\cite{Balazs:2007hr}.
The function $h_{g}^{(1)}(\theta_{*})$ denotes an ${\cal O}(\alpha_{s}/\pi)$
correction to the hard-scattering contribution ${\cal H}$ in the
resummed cross section, cf. Sec.~\ref{subsection:CompleteResummed}.
The convolutions $\left[P_{g/a}\otimes f_{a/h}\right]$ and $\left[{\mathcal{C}}_{g/a}^{(1,c)}\otimes f_{a/h}\right]$,
defined for two functions $f(x,\mu_{F})$ and $g(x,\mu_{F})$ as\[
[f\otimes g](x,\mu_{F})\equiv\int_{x}^{1}\frac{d\xi}{\xi}f(\xi,\mu_{F})g(\frac{x}{\xi},\mu_{F}),\]
 are summed over the intermediate parton's flavors $a=g,\, q_{S}$
(gluon and the flavor-singlet combination of quark-scattering channels).
In addition to the conventional splitting functions $P_{g/g}(x)$
and $P_{g/q_{S}}(x)$ arising in $F_{g,+}$, a new splitting function\begin{equation}
P_{g/g}^{\prime}(x)=2C_{A}(1-x)/x,\label{Pggprime}\end{equation}
 where $C_{A}=N_{c}=3,$ enters the $\varphi_{*}$-dependent part
of the asymptotic cross section through $F_{g,+}^{\prime}.$

For completeness, the small-$Q_{T}$ asymptotic form Eq.~(\ref{ASYgg})
for the $gg+gq_{S}$ channels is derived in Appendix~\ref{Appendix:ASYgg}.
The existence of the $\varphi_{*}-$dependent singular contribution
proportional to $\Sigma_{g}^{\prime}(\theta_{*},\varphi_{*})$ is
established by examining the factorization of the $2\rightarrow3$
cross section in the limit of a collinear gluon emission. It follows
directly from factorization rules for helicity amplitudes \cite{Bern:1993mq,Bern:1993qk,Bern:1994zx,Bern:1994fz,Bern:1998sc,Bern:1999ry,Kosower:1999rx},
as well as from the dipole factorization formalism \cite{Catani:1996vz}.

In contrast, the NLO quark-antiquark contribution $q\bar{q}\rightarrow\gamma\gamma$
does not include a spin-flip contribution, as a result of the simple
structure of the Born contribution in $q\bar{q}$ scattering (see
also Sec.~\ref{subsection:Spin-Flip}).

\subsection{Resummation \label{subsection:ResummationOne}}

To predict the shape of $d\sigma/dQ_{T}$ distributions, we perform
an all-orders summation of singularities $\delta(\vec{Q}_{T})$ and
$\left[Q_{T}^{-2}\ln^{p}\left(Q^{2}/Q_{T}^{2}\right)\right]_{+}$
in the asymptotic cross section, which coincides with the perturbative
expansion of the resummed small-$Q_{T}$ cross section obtained within
the Collins-Soper-Sterman formalism \cite{Collins:1981uk,Collins:1981va,Collins:1984kg}.
In this formalism, we write the fully differential cross section as
\begin{eqnarray}
\frac{d\sigma(h_{1}h_{2}\rightarrow\gamma\gamma)}{dQ\, dQ_{T}^{2}\, dy\, d\Omega_{*}}=W(Q,Q_{T},y,\Omega_{*})+Y(Q,Q_{T},y,\Omega_{*}).\label{FullResum}\end{eqnarray}
 The term $W$ contains large logarithmic contributions of the form
$\ln^{p}(Q/Q_{T})$ from initial-state radiation, while $Y$ is free
of these logs and calculated using collinear QCD factorization (cf.
the end of Sec.~\ref{subsection:CompleteResummed}).

The function $W$ may be expressed as a Fourier-Bessel transform of
a function $\widetilde{W}(Q,b,y,\Omega_{*})$ in the impact parameter
(${\vec{b}}$) space, \begin{equation}
W(Q,Q_{T},y,\Omega_{*})=\int\frac{d\vec{b}}{(2\pi)^{2}}e^{i\vec{Q}_{T}\cdot\vec{b}}\widetilde{W}(Q,b,y,\Omega_{*}).\label{FourierIntegral}\end{equation}
 The generic form of $\widetilde{W}(Q,b,y,\Omega_{*})$ in the $q\bar{q}+qg\rightarrow\gamma\gamma$
and $gg+gq_{S}\rightarrow\gamma\gamma$ channels can be determined
by solving evolution equations for the gauge- and renormalization-group
invariance of $\widetilde{W}(Q,b,y,\Omega_{*})$:\begin{eqnarray}
\widetilde{W}(Q,b,y,\Omega_{*}) & = & \sum_{a}\sum_{\lambda_{1},\lambda_{1}^{\prime},\lambda_{2},\lambda_{2}^{\prime},\lambda_{3},\lambda_{4}}{\mathcal{H}}_{a}^{\lambda_{1}\lambda_{2}\lambda_{3}\lambda_{4}}(Q,\Omega_{*})\left({\mathcal{H}}_{a}^{\lambda_{1}^{\prime}\lambda_{2}^{\prime}\lambda_{3}\lambda_{4}}(Q,\Omega_{*})\right)^{*}\nonumber \\
 &  & \hspace{84pt}\times{\mathcal{P}}_{a/h_{1}}^{\lambda_{1}\lambda_{1}^{\prime}}(x_{1},\vec{b})\,{\mathcal{P}}_{\bar{a}/h_{2}}^{\lambda_{2}\lambda_{2}^{\prime}}(x_{2},\vec{b})\, e^{-{\mathcal{S}}_{a}(Q,b)}.\label{Eq:Wt}\end{eqnarray}
 It is composed of the hard-scattering function ${\mathcal{H}}_{a}^{\lambda_{1}\lambda_{2}\lambda_{3}\lambda_{4}}(Q,\Omega_{*})$
and its complex conjugate, $\left({\mathcal{H}}_{a}^{\lambda_{1}^{\prime}\lambda_{2}^{\prime}\lambda_{3}\lambda_{4}}(Q,\Omega_{*})\right)^{*};$
the Sudakov exponential $\exp\left(-{\mathcal{S}}_{a}(Q,b)\right)$;
and parton distribution matrices ${\mathcal{P}}_{a/h_{i}}^{\lambda_{i}\lambda_{i}^{\prime}}(x_{i},\vec{b})$.

\begin{figure}
\begin{centering}
\includegraphics[width=8cm,keepaspectratio]{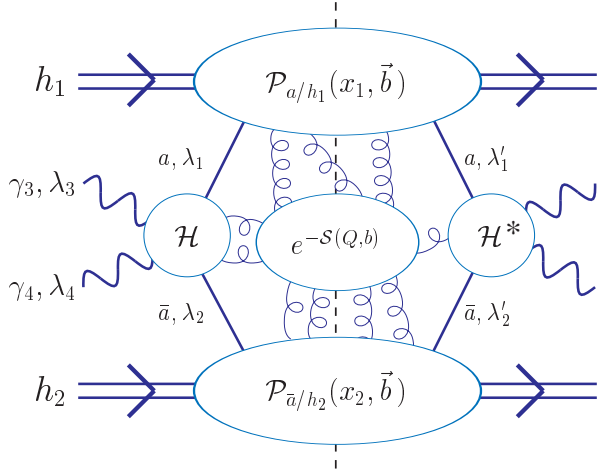}
\par\end{centering}

\caption{The structure of the resummed form factor $\widetilde{W}(Q,b,y,\Omega_{*})$.
\label{Fig:Factorization}}
\end{figure}

The multiplicative structure of Eq.~(\ref{Eq:Wt}) reflects the topology
of the dominant cut diagrams in the small-$Q_{T}$ cross sections
shown in Fig.~\ref{Fig:Factorization}. The function ${\mathcal{H}}_{a}^{\lambda_{1}\lambda_{2}\lambda_{3}\lambda_{4}}$
describes the hard $2\rightarrow2$ scattering subprocess $a(p_{1},\lambda_{1})+\bar{a}(p_{2},\lambda_{2})\rightarrow\gamma(p_{3},\lambda_{3})+\gamma(p_{4},\lambda_{4})$,
with $a=u,\bar{u},d,\bar{d},...$ in $q\bar{q}\rightarrow\gamma\gamma$,
and $a=\bar{a}=g$ in $gg\rightarrow\gamma\gamma$. All momenta in
${\mathcal{H}}$ have virtualities of order $Q^{2}.$ For now, we
consider the leading contribution to ${\mathcal{H}}_{a}^{\lambda_{1}\lambda_{2}\lambda_{3}\lambda_{4}},$
which reads as ${\mathcal{H}}_{a}^{\lambda_{1}\lambda_{2}\lambda_{3}\lambda_{4}}=\sqrt{\sigma_{a}^{(0)}}{\mathcal{M}}_{4}(\lambda_{1},\lambda_{2},\lambda_{3},\lambda_{4}),$
where the Born helicity amplitude ${\mathcal{M}}_{4}(\lambda_{1},\lambda_{2},\lambda_{3},\lambda_{4})$
and overall constant normalization $\sigma_{a}^{(0)}$ are introduced
in Sec.~\ref{subsection:AsymptoticISR}. Sometimes ${\mathcal{H}}_{a}^{\lambda_{1}\lambda_{2}\lambda_{3}\lambda_{4}}$
also includes finite parts of higher-order $2\rightarrow2$ virtual
corrections, as discussed in Sec. \ref{subsection:CompleteResummed}.

Similarly, $\left({\mathcal{H}}_{a}^{\lambda_{1}^{\prime}\lambda_{2}^{\prime}\lambda_{3}\lambda_{4}}(Q,\Omega_{*})\right)^{*}$
arises from the complex-conjugate amplitude ${\mathcal{M}}_{4}^{*}(\lambda_{1},\lambda_{2},\lambda_{3},\lambda_{4})$
and possible loop corrections to it. The helicities $\lambda_{1}^{\prime}$
and $\lambda_{2}^{\prime}$ in $\left({\mathcal{H}}_{a}^{\lambda_{1}^{\prime}\lambda_{2}^{\prime}\lambda_{3}\lambda_{4}}\right)^{*}$
need not coincide with $\lambda_{1}$ and $\lambda_{2}$ in ${\mathcal{H}}_{a}^{\lambda_{1}\lambda_{2}\lambda_{3}\lambda_{4}}$
. The right-hand side of Eq.~(\ref{Eq:Wt}) is summed over flavors
$a$ and helicities $\lambda_{k},\lambda_{k}^{\prime}$ of the partons
entering ${\mathcal{H}}{\mathcal{{H}^{*}}}$, as well as over helicities
$\lambda_{3}$ and $\lambda_{4}$ of the final-state photons.

The Sudakov exponent \begin{eqnarray}
{\mathcal{S}}_{a}(Q,b)=\int_{C_{1}^{2}/b^{2}}^{C_{2}Q^{2}}\frac{d\bar{\mu}^{2}}{\bar{\mu}^{2}}\left[{\mathcal{A}}_{a}\left(C_{1},\bar{\mu}\right)\ln\left(\frac{C_{2}^{2}Q^{2}}{\bar{\mu}^{2}}\right)+{\mathcal{B}}_{a}\left(C_{1},C_{2},\bar{\mu}\right)\right]\label{Sudakov}\end{eqnarray}
 resums contributions from the initial-state soft and soft-collinear
gluon emission (indicated by gluon lines connecting $e^{-{\cal S}}$
to ${\cal H}$, ${\cal H}^{*},$ and ${\cal P}_{a/h}(x,\vec{b})$
in Fig.~\ref{Fig:Factorization}). Here $C_{1}$ and $C_{2}$ are
constants of order unity. The functions ${\mathcal{A}}_{a}\left(C_{1},\bar{\mu}\right)$
and ${\mathcal{B}}_{a}\left(C_{1},C_{2},\bar{\mu}\right)$ can be
evaluated in perturbation theory at large scales $\bar{\mu}^{2}\gg\Lambda_{QCD}^{2}$,
hence for large $Q$ and small $b$.

The collinear emissions are described by parton distribution matrices
${\mathcal{P}}_{a/h}^{\lambda\lambda^{\prime}}(x,\vec{b})$, where
$\lambda$ and $\lambda^{\prime}$ denote the helicity state of the
intermediate parton $a$ to the left and right of the unitarity cut
in Fig.~\ref{Fig:Factorization}. The matrix ${\mathcal{P}}_{a/h}^{\lambda\lambda^{\prime}}(x,\vec{b})$
is derived from a matrix element of the light-cone correlator \cite{Ralston:1979ys,Soper:1976jc,Soper:1979fq,Ali:1992qj,Bashinsky:1998if}
for finding parton $a$ inside the parent hadron $h$.

It is convenient to introduce sums of diagonal and off-diagonal entries
of the helicity matrix ${\mathcal{P}}_{a/h}^{\lambda\lambda^{\prime}}(x,\vec{b})$,\begin{equation}
{\mathcal{P}}_{a/h}(x,\vec{b})=\sum_{\lambda}{\mathcal{P}}_{a/h}^{\lambda\lambda}(x,\vec{b}),\end{equation}
 and \begin{equation}
{\mathcal{P}}_{a/h}^{\prime}(x,\vec{b})=\sum_{\lambda}{\mathcal{P}}_{a/h}^{\lambda,-\lambda}(x,\vec{b}).\end{equation}
 In this notation, Eq.~(\ref{Eq:Wt}) can be rewritten as \begin{eqnarray}
\widetilde{W}(Q,b,y,\Omega_{*}) & = & \frac{1}{S}e^{-{\mathcal{S}}_{a}(Q,b)}\sum_{a}\Biggl\{\Sigma_{a}(\theta_{*}){\mathcal{P}}_{a/h_{1}}(x_{1},\vec{b})\,{\mathcal{P}}_{\bar{a}/h_{2}}(x_{2},\vec{b})+\nonumber \\
 & + & \Sigma_{a}^{\prime}(\theta_{*},\varphi_{*})\left[{\mathcal{P}}_{a/h_{1}}^{\prime}(x_{1},\vec{b})\,{\mathcal{P}}_{\bar{a}/h_{2}}(x_{2},\vec{b})+{\mathcal{P}}_{a/h_{1}}(x_{1},\vec{b})\,{\mathcal{P}}_{\bar{a}/h_{2}}^{\prime}(x_{2},\vec{b})\right]\label{Eq:Wt2}\\
 & + & \Sigma_{a}^{\prime\prime}(\theta_{*},\varphi_{*}){\mathcal{P}}_{a/h_{1}}^{\prime}(x_{1},\vec{b})\,{\mathcal{P}}_{\bar{a}/h_{2}}^{\prime}(x_{2},\vec{b})\Biggr\},\nonumber \end{eqnarray}
 where \begin{eqnarray}
\Sigma_{a}(\theta_{*}) & \equiv & \sum_{\lambda_{1},\lambda_{2},\lambda_{3},\lambda_{4}}\left|{\mathcal{H}}_{a}^{\lambda_{1}\lambda_{2}\lambda_{3}\lambda_{4}}\right|^{2},\\
\Sigma_{a}^{\prime}(\theta_{*},\varphi_{*}) & \equiv & \sum_{\lambda_{1},\lambda_{2},\lambda_{3},\lambda_{4}}{\mathcal{H}}_{a}^{\lambda_{1}\lambda_{2}\lambda_{3}\lambda_{4}}\left({\mathcal{H}}_{a}^{-\lambda_{1}\lambda_{2}\lambda_{3}\lambda_{4}}\right)^{*},\end{eqnarray}
 and \begin{equation}
\Sigma_{a}^{\prime\prime}(\theta_{*},\varphi_{*})\equiv\sum_{\lambda_{1},\lambda_{2},\lambda_{3},\lambda_{4}}{\mathcal{H}}_{a}^{\lambda_{1}\lambda_{2}\lambda_{3}\lambda_{4}}\left({\mathcal{H}}_{a}^{-\lambda_{1}-\lambda_{2}\lambda_{3}\lambda_{4}}\right)^{*}.\end{equation}

The unpolarized parton distribution ${\mathcal{P}}_{a/h}(x,\vec{b})$
coincides with the Fourier-Bessel transform of the unpolarized transverse-momentum-dependent
(TMD) parton density ${\mathcal{P}}_{a/h}(x,\vec{k}_{T})$ \cite{Collins:1981uw}
for finding parton $a$ with light-cone momentum fraction $x$ and
transverse momentum $\vec{k}_{T}$. At small $b,$ ${\mathcal{P}}_{a/h}(x,\vec{b})$
is reduced to a convolution of unpolarized $k_{T}-$integrated parton
densities $f_{a/h}(x,\mu)$ and Wilson coefficient functions ${\mathcal{C}}_{a/a^{\prime}}(x,b;C/C_{2},\mu)$,
evaluated at a factorization scale $\mu$ of order $1/b$:\begin{eqnarray}
\left.{\mathcal{P}}_{a/h}(x,\vec{b})\right|_{b^{2}\ll\Lambda_{QCD}^{-2}} & = & \sum_{a^{\prime}}\left[\int_{x}^{1}{\frac{d\xi}{\xi}}{\mathcal{C}}_{a/a^{\prime}}\left(\frac{x}{\xi},b;\frac{C_{1}}{C_{2}},\mu\right)f_{a^{\prime}/h}(\xi,\mu)\right].\label{Eq:CalP}\end{eqnarray}
 Perturbative entries with $\lambda_{i}=\lambda_{i}^{\prime}$ reduce
in total to the product of the unpolarized Born scattering probability
and unpolarized resummed functions:\begin{eqnarray}
\left.\widetilde{W}(Q,b,y,\Omega_{*})\right|_{\lambda_{i}=\lambda_{i}^{\prime}} & = & \sum_{a}\frac{\Sigma_{a}(\theta_{*})}{S}e^{-{\mathcal{S}}_{a}(Q,b)}\nonumber \\
 & \times & \left[{\mathcal{C}}_{a/c_{1}}\otimes f_{c_{1}/h_{1}}\right](x_{1},b;\mu)\left[{\mathcal{C}}_{\bar{a}/c_{2}}\otimes f_{c_{2}/h_{2}}\right](x_{2},b;\mu).\label{WSpinDiagonal}\end{eqnarray}
 The function $\Sigma_{g}(\theta_{*})$ is shown explicitly in Eq.~(\ref{Sigmag}).

\subsection{Spin-flip term in gluon scattering \label{subsection:Spin-Flip}}

We concentrate in this subsection on the spin-flip distribution ${\mathcal{P}}_{g/h}^{\prime}(x,\vec{b})$
in gluon scattering. Its existence is warranted by basic symmetries
of helicity- and transverse-momentum-dependent gluon distribution
functions \cite{Mulders:2000sh}. This function, which describes interference
of the amplitudes for nearly collinear gluons with opposite helicities,
coincides with the function $H^{\perp}$ in Ref.~\cite{Mulders:2000sh}
up to an overall factor. It contributes to \emph{unpolarized} $Q_{T}$
distributions, because the hard-scattering product ${\mathcal{H}}_{g}^{\lambda_{1}\lambda_{2}\lambda_{3}\lambda_{4}}\left({\mathcal{H}}_{g}^{\lambda_{1}^{\prime}\lambda_{2}^{\prime}\lambda_{3}\lambda_{4}}\right)^{*}$
(with $ $ ${\mathcal{H}}_{g}^{\lambda_{1}\lambda_{2}\lambda_{3}\lambda_{4}}$
given by the quark box helicity amplitude in Fig.~\ref{Fig:FeynDiag}(a))
does not vanish for $\lambda_{1}=-\lambda_{1}^{\prime}$ or $\lambda_{2}=-\lambda_{2}^{\prime}$
. The presence of ${\mathcal{P}}_{g/h}^{\prime}(x,\vec{b})$ modifies
dependence of the resummed cross section on the photon's azimuthal
angle $\varphi_{*}$ in the Collins-Soper frame. It vanishes after
the integration over $\varphi_{*}$ is performed. In contrast, the
helicity-diagonal part of $\widetilde{W}(Q,b,y,\Omega_{*})$ is independent
of $\varphi_{*}$, cf. Eq.~(\ref{WSpinDiagonal}).

The gluon function ${\mathcal{P}}_{g/h}^{\prime}(x,\vec{b})$ is invariant
under time reversal (i.e., is $T$-even) and acquires large contributions
proportional to the unpolarized $T$-even PDF's ${\mathcal{P}}_{g/h}(x,\vec{b})$
in the process of gluon radiation. These contributions require resummation
via PDF evolution equations (similar to Dokshitzer-Gribov-Lipatov-Altarelli-Parisi
equations \cite{Dokshitzer:1977sg,Gribov:1972ri,Gribov:1972rt,Altarelli:1977zs})
in order to predict the $\varphi_{*}$ dependence in the $gg$ channel.

\begin{figure}
\includegraphics[width=0.8\textwidth,keepaspectratio]{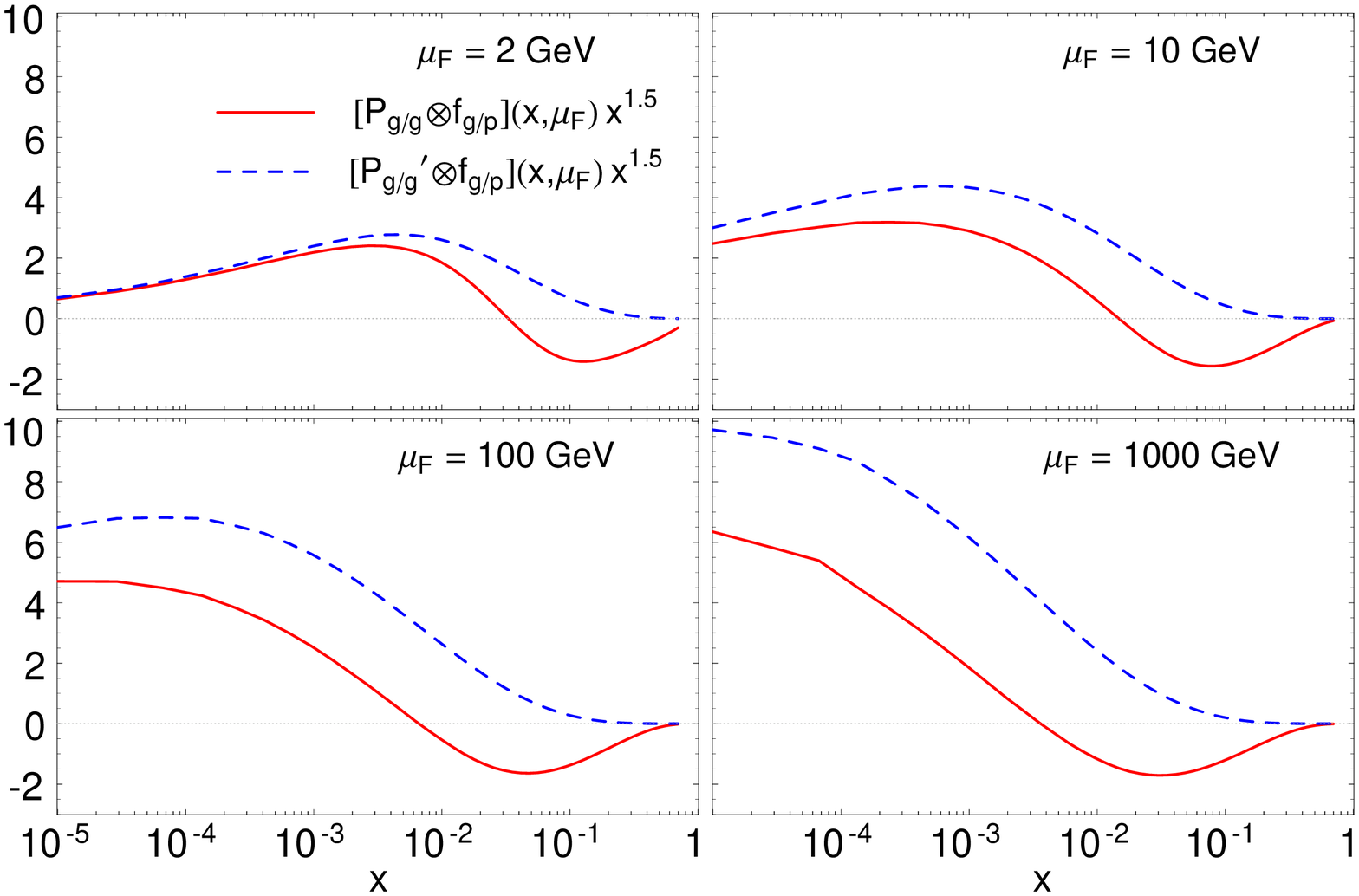}

\caption{Comparison of $[P_{g/g}\otimes f_{g/p}](x,\mu_{F})$ and $[P_{g/g}^{\prime}\otimes f_{g/p}](x,\mu_{F})$
for the gluon PDF $f_{g/p}(x,\mu_{F})$ in the proton (multiplied
by $x^{1.5}$ to better illustrate the small-$x$ region) at several
values of the factorization scale $\mu_{F}.$ \label{fig:PggxQ}}
\end{figure}

At one loop, the mixing of spin-flip and unpolarized gluon PDF's is
driven by the convolution $\left[P_{g/g}^{\prime}\otimes f_{g/h}\right](x,\mu_{F})$
of the spin-flip splitting function $P_{g/g}^{\prime}(x)$ shown in
Eq.~(\ref{Pggprime}) with the gluon PDF $f_{g/h}(x,\mu_{F})$. This
convolution may be comparable to or exceed the analogous convolution
$\left[P_{g/g}\otimes f_{g/h}\right](x,\mu_{F})$ of the unpolarized
splitting function $P_{g/g}(x)$ for some $x$ and $\mu_{F}$ values,
as shown in Fig.~\ref{fig:PggxQ}. As a result of the mixing, an
additional $\varphi_{*}$-dependent term \begin{eqnarray}
 &  & \frac{\Sigma_{g}^{\prime}(\theta_{*},\varphi_{*})}{2\pi SQ_{T}^{2}}\frac{\alpha_{s}}{\pi}\Bigl(\left[P_{g/g}^{\prime}\otimes f_{g/h_{1}}\right](x_{1},\mu_{F})\, f_{g/h_{2}}(x_{2},\mu_{F})\nonumber \\
 &  & \hspace{77pt}+f_{g/h_{1}}(x_{1},\mu_{F})\left[P_{g/g}^{\prime}\otimes f_{g/h_{2}}\right](x_{2},\mu_{F})\Bigr)\label{ASYprime}\end{eqnarray}
 arises in the unpolarized ${\mathcal{O}}(\alpha_{s})$ asymptotic
piece, cf. Eq.~(\ref{ASYgg}). It is produced by the perturbative
expansion of the entry proportional to $\Sigma_{g}^{\prime}(\theta_{*},\varphi_{*}){\mathcal{P}}_{g/h}(x_{i},\vec{b}){\mathcal{P}}_{g/h}^{\prime}(x_{j},\vec{b})$
in $\widetilde{W}(Q,b,y,\Omega_{*})$, with $\Sigma_{g}^{\prime}(\theta_{*},\varphi_{*})$
shown explicitly in Eqs.~(\ref{Sigmagprime}) and (\ref{Lprime}).
Generally, the $\varphi_{*}$-dependent contribution is not small,
even though it is suppressed comparatively to the unpolarized collinear
contribution by the ratio $L_{g}^{\prime}(\theta_{*})/L_{g}(\theta_{*})$
shown in Fig.~\ref{Fig:LLprime}. For example, for $Q=100$ GeV at
the LHC, its magnitude constitutes up to about a half of the collinear
unpolarized asymptotic contribution,\begin{equation}
\frac{\Sigma_{g}(\theta_{*})}{2\pi SQ_{T}^{2}}\frac{\alpha_{s}}{\pi}\Bigl(\left[P_{g/a}\otimes f_{a/h_{1}}\right](x_{1},\mu_{F})\, f_{g/h_{2}}(x_{2},\mu_{F})+f_{g/h_{1}}(x_{1},\mu_{F})\left[P_{g/a}\otimes f_{a/h_{2}}\right](x_{2},\mu_{F})\Bigr).\label{ASYNoSpinFlip}\end{equation}
 The ${\mathcal{O}}(\alpha_{s})$ spin-flip $gg$ contribution does
not mix with the $gq_{S}$ contribution.

In terms of the reduced matrix elements $M_{\lambda_{1}\lambda_{2}\lambda_{3}\lambda_{4}}^{(1)}$
defined in \cite{Bern:2001df}, the double spin-flip hard vertex function
$\Sigma_{g}^{\prime\prime}(\theta_{*},\varphi_{*})$ is\begin{eqnarray}
\Sigma_{g}^{\prime\prime}(\theta_{*},\varphi_{*}) & = & \sigma_{g}^{(0)}\left(L_{1g}^{\prime\prime}(\theta_{*})+L_{2g}^{\prime\prime}(\theta_{*})\cos\left(4\varphi_{*}\right)\right),\end{eqnarray}
 where\begin{eqnarray}
L_{1g}^{\prime\prime}(\theta_{*}) & = & 4{\mbox{Re}}\left(M_{1,1,-1,-1}^{(1)}+1\right),\end{eqnarray}
 and\begin{eqnarray}
L_{2g}^{\prime\prime}(\theta_{*}) & = & 4{\mbox{Re}}\left(M_{1,-1,-1,1}^{(1)}M_{1,-1,1,-1}^{(1)*}+1\right).\end{eqnarray}
 The perturbative expansion of the resummed entry proportional to
$\Sigma_{g}^{\prime\prime}(\theta_{*},\varphi_{*}){P}_{g/h}^{\prime}(x_{i},\vec{b}){P}_{g/h}^{\prime}(x_{j},\vec{b})$
produces an NNLO term in the unpolarized $gg$ asymptotic piece, \begin{equation}
\frac{\Sigma_{g}^{\prime\prime}(\theta_{*},\varphi_{*})}{2\pi SQ_{T}^{2}}\frac{\alpha_{s}^{2}}{\pi^{2}}\left[P_{g/g}^{\prime}\otimes f_{g/h_{1}}\right](x_{1},\mu_{F})\left[P_{g/g}^{\prime}\otimes f_{g/h_{2}}\right](x_{2},\mu_{F}).\end{equation}

The analogous quark function ${\mathcal{P}}_{q_{i}/h}^{\prime}(x,\vec{k}_{T})$
corresponds to the transversity distribution \cite{Tangerman:1994eh}
and is odd under time reversal ($T$-odd). It cannot be generated
radiatively through conventional PDF evolution from the $T$-even
unpolarized function ${\mathcal{P}}_{q_{i}/h}(x,\vec{k}_{T})$ and
does not contribute to the NLO asymptotic term. We find $\Sigma_{q}^{\prime}=0,$
because the non-vanishing amplitudes ${\cal H}_{q}(q_{1}^{\lambda_{1}},\bar{q}_{2}^{\lambda_{2}},\gamma_{3}^{\lambda_{3}},\gamma_{4}^{\lambda_{4}})$
must have opposite helicities of the quark and antiquark ($\lambda_{1}=-\lambda_{2}$).
Therefore, the functions ${\mathcal{P}}_{q_{i}/h}^{\prime}(x,\vec{b})$
contribute in pairs through the term proportional to $\Sigma_{q}^{\prime\prime}(\varphi_{*})=-(\alpha^{2}e_{i}^{4}\pi/(2N_{c}Q^{2}))\,\cos2\varphi_{*}.$
These contributions are anticipated to be much smaller than the usual
spin-average contribution and negligible at large $Q$, in analogy
to unpolarized Drell-Yan production \cite{Henneman:2001ev,Boer:2001he}.

In summary, the azimuthal angle ($\varphi_{*}$) dependence of photons
in the $gg$ scattering channel is affected by large QCD contributions
associated with interference between gluons of opposite helicities.
These logarithmic corrections may arise at NLO through QCD radiation
from conventional unpolarized PDF's, a mechanism that is unique to
gluon scattering. Other types of spin-interference contributions (not
considered here) involve spin-flip PDF's only. The soft and collinear
logarithms associated with the spin-flip contributions must be resummed
along the lines discussed in Ref.~\cite{Idilbi:2004vb}. Given that
$gg\rightarrow\gamma\gamma$ is the subleading production channel
at the Tevatron and at the LHC, we henceforth neglect the gluon spin-flip
contributions to the resummed $\widetilde{W}(Q,b,y,\Omega_{*})$,
while subtracting the corresponding $\varphi_{*}$-dependent asymptotic
contribution from the finite-order $2\rightarrow3$ cross section.
The nature of $gg$ spin-flip contributions can be explored by measuring
the double-differential distribution in $\varphi_{*}$ and $Q_{T}$
at the LHC, a topic that is interesting also from the point of view
of the Higgs boson search. Full resummation of the gluon spin-flip
contributions may be needed in the future.

\subsection{Complete expressions for resummed cross sections \label{subsection:CompleteResummed}}

In this Section, we review complete expressions for the unpolarized
resummed cross sections, starting from the perturbative QCD approximation
$\widetilde{W}_{pert}(Q,b,y,\Omega_{*})$ valid at small impact parameters
$b^{2}\ll1\mbox{ GeV}^{-2}$. For a hard-scattering function $\sum_{hel.}\left|{\mathcal{H}}(Q,\theta_{*})\right|^{2}\equiv\Sigma_{a}(\theta_{*})h_{a}^{2}(Q,\theta_{*}),$
the form factor $\widetilde{W}_{pert}(Q,b,y,\Omega_{*})$ is \begin{eqnarray}
\widetilde{W}_{pert}(Q,b,y,\theta_{*}) & = & \sum_{a}\frac{\Sigma_{a}(\theta_{*})}{S}h_{a}^{2}(Q,\theta_{*})e^{-{\mathcal{S}}_{a}(Q,b)}\nonumber \\
 & \times & \left[{\mathcal{C}}_{a/c_{1}}\otimes f_{c_{1}/h_{1}}\right](x_{1},b;\mu)\left[{\mathcal{C}}_{\bar{a}/c_{2}}\otimes f_{c_{2}/h_{2}}\right](x_{2},b;\mu).\label{UnpW2}\end{eqnarray}
 The Sudakov function is defined in Eq.~(\ref{Sudakov}), and the
function $h_{a}(Q,\theta_{*})$ collects radiative contributions to
${\mathcal{H}}(Q,\theta_{*})$ arising at NLO and beyond. We compute
the functions $h_{a},$ ${\mathcal{A}}_{a}$, ${\mathcal{B}}_{a}$
and ${\mathcal{C}}_{a/c}$ up to orders $\alpha_{s},$ $\alpha_{s}^{3},$
$\alpha_{s}^{2},$ and $\alpha_{s},$ respectively. The ${\mathcal{A}}_{a}$,
${\mathcal{B}}_{a}$ and ${\mathcal{C}}_{a/c}$ coefficients are taken
from Refs.~\cite{Balazs:1997hv,Nadolsky:2002gj,Yuan:1991we,Vogt:2004mw,deFlorian:2000pr,Moch:2004pa}
and listed in a consistent notation in Ref.~\cite{Balazs:2007hr}.

We use a procedure outlined in Ref.~\cite{Balazs:1997xd} to join
the small-$Q_{T}$ resummed cross sections $W$ with the large-$Q_{T}$
NLO cross sections $P$. In Eq.~(\ref{FullResum}), $Y\equiv P-A$
is the difference between the perturbative cross section $P$ and
its small-$Q_{T}$ asymptotic expansion $A$, explicitly given in
Eq.~(\ref{ASYgg}). For each value of $Q$ and $y$ of the $\gamma\gamma$
pair, $W+Y$ approaches $P$ from above and eventually becomes smaller
than $P$ as $Q_{T}$ increases. We use $W+Y$ as our prediction at
$Q_{T}$ values below this point of crossing and the finite-order
cross section $P$ at $Q_{T}$ above the crossing point.

The final cross sections depend on several factorization scales: $C_{1}/b$,
$C_{2}Q$, $\mu\equiv C_{3}/b$ in the $W$ term, and $\mu_{F}\equiv C_{4}Q$
in the $Y$ term. Here $C_{i}$ ($i=1,..4$) are dimensionless constants
of order unity, chosen as $C_{2}=C_{4}=1$, $C_{1}=C_{3}=2e^{-\gamma_{E}}=1.123...$
by default. These choices simplify perturbative coefficients by eliminating
scale-dependent logarithmic terms, cf. the appendix in Ref.~\cite{Balazs:2007hr}.
Dependence on the scale choice is studied in Section~\ref{Sec:Phenomenology}.

In the general formulation of CSS resummation presented in \cite{Collins:1981uk,Collins:1981va},
one has the freedom to choose different {}``resummation schemes'',
resulting effectively in variations in the form of $h_{a}(Q,\theta_{*})$.
These differences are compensated, up to higher-order corrections,
by adjustments in the functions ${\mathcal{B}}$ and ${\mathcal{C}}$.

In {}``the CSS resummation scheme'' \cite{Collins:1984kg}, one
chooses $h_{a}(Q,\theta_{*})=1,$ while including the virtual corrections
to the $2\rightarrow2$ scattering process in ${\mathcal{B}}$ and
${\mathcal{C}}.$ In this scheme, some ${\mathcal{B}}$ and ${\mathcal{C}}$
coefficients depend on the $2\rightarrow2$ hard scattering process
and also on $\theta_{*}$.

In an alternative prescription by Catani, de Florian and Grazzini
\cite{Catani:2000vq}, {}``the CFG resummation scheme'', one keeps
the $2\rightarrow2$ virtual corrections within a single function
$\left|{\mathcal{H}}(Q,\theta_{*})\right|^{2}.$ In this case, the
${\mathcal{B}}$ and ${\mathcal{C}}$ functions depend only on the
initial state. Most of our numerical calculations are realized in
the CSS resummation scheme, with a few made in the CFG scheme for
comparison purposes.

In impact parameter ($b$) space used in the resummation procedure,
we must integrate into the nonperturbative region of large $b$, cf.~Eq.~(\ref{FourierIntegral}).
Contributions from this region are known to be suppressed at high
energies~\cite{Berger:2002ut}, but some residual dependence may
remain. In the $q\bar{q}+qg\rightarrow\gamma\gamma$ channel, our
model for the nonperturbative contributions (denoted as KN1 \cite{Konychev:2005iy})
is derived from the analysis of Drell-Yan pair and $Z$ boson production.
The nonperturbative function in this model is dominated at large $Q$
by a soft contribution, which does not depend on the flavor of initial-state
light quarks. This function is therefore expected to be applicable
to the $q\bar{q}+qg\rightarrow\gamma\gamma$ process.

The nonperturbative function in the $gg+gq_{S}$ channel, which is
yet to be measured directly, is approximated by the nonperturbative
function for the $q\bar{q}+gq$ channel multiplied by the ratio $C_{A}/C_{F}=9/4$
of the color factors $C_{A}$ and $C_{F}$ for the leading soft contributions
in the $gg$ and $q\bar{q}$ channels. This ansatz suggests stronger
dependence of the $gg+gq_{S}$ channel on the nonperturbative input
compared to the $q\bar{q}+qg$ channels. It leads to small differences
from the prescription used in Refs.~\cite{Balazs:1997hv,Balazs:1999yf},
where only the leading $\ln Q$ term of the nonperturbative function
was rescaled. To examine the dependence of the resummed cross sections
on the nonperturbative model, we evaluate some of them assuming an
alternative (BLNY) parameterization of the nonperturbative function
\cite{Landry:2002ix}.

\section{Numerical Results \label{Sec:Phenomenology}}

The analytical results of Sec.~III are implemented in our computer
codes \textsc{Legacy} and \textsc{ResBos} \cite{Ladinsky:1993zn,Landry:2002ix,Balazs:1997xd,Balazs:1999gh}.
We use the same parameters as in the calculation of Ref.~\cite{Balazs:2006cc},
and we concentrate on the region $Q_{T}<Q$ where our calculation
is most reliable \cite{Balazs:2006cc}.

\begin{figure*}
\includegraphics[width=100cm,height=9.5cm,keepaspectratio]{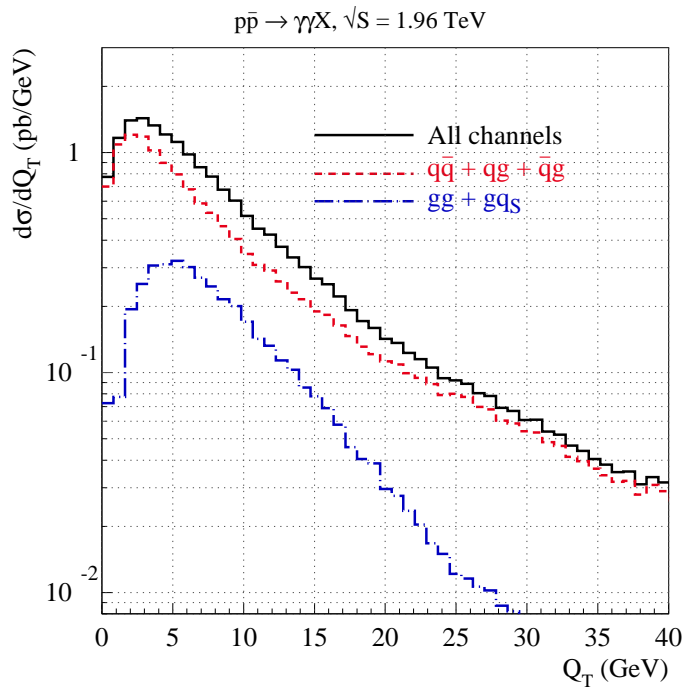}

\caption{Parton flavor decomposition of the resummed transverse momentum distribution
at the energy $\sqrt{S}=1.96$ TeV of the Tevatron Run-2. The total
(solid), $q{\bar{q}}+qg$ (dashes), and $gg+gq_{S}$ (dash-dots) initial-state
contributions are shown separately. \label{Fig:bStarCDF}}
\end{figure*}

\subsection{Results for Run 2 at the Tevatron}

In this section, we present our results for the Tevatron $p\bar{p}$
collider at $\sqrt{S}=1.96$~TeV. We make the same restrictions on
the final-state photons as those used in the experimental measurement
by the Collider Detector at Fermilab (CDF) collaboration~\cite{Acosta:2004sn}:
transverse momentum $p_{T}^{\gamma}>p_{T\, min}^{\gamma}=14~(13)$~GeV
for the harder (softer) photon, and rapidity $|y^{\gamma}|<0.9$ for
each photon. We impose photon isolation by requiring the hadronic
transverse energy not to exceed $1$ GeV in the cone $\Delta R=0.4$
around each photon, as specified in the CDF publication. We also require
the angular separation $\Delta R_{\gamma\gamma}$ between the photons
to be larger than 0.3.

We focus in this paper on the role of the $gg$ contribution, referring
to our other papers \cite{Balazs:2006cc,Balazs:2007hr} for a more
complete treatment.

\begin{figure*}
\includegraphics[width=0.7\textwidth,keepaspectratio]{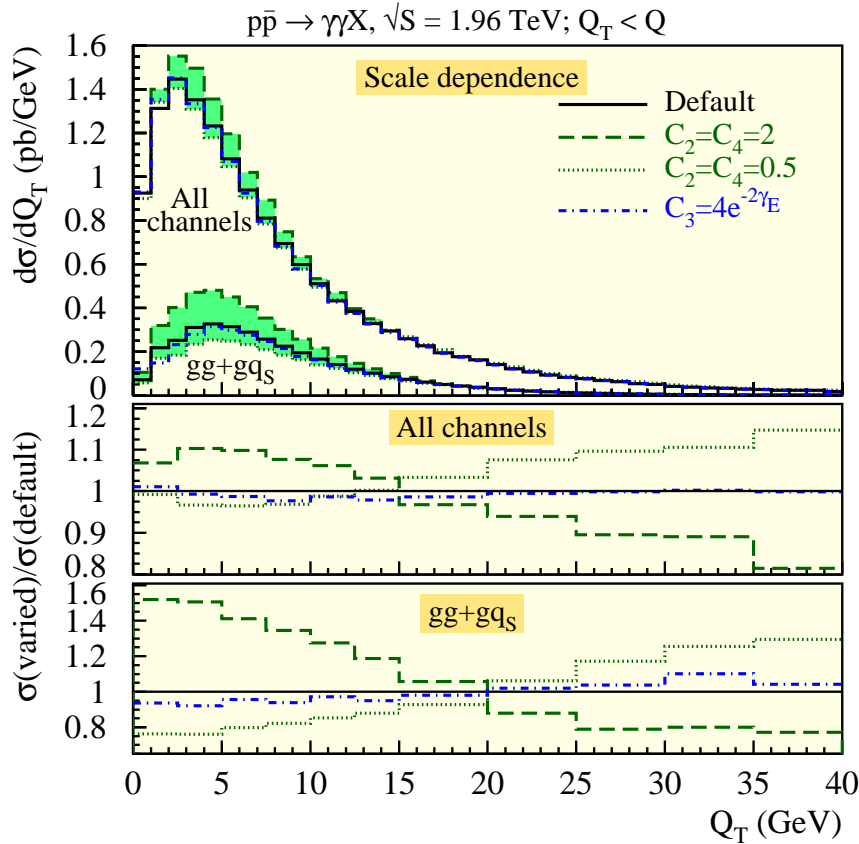}

\caption{Scale dependence of the cross sections in $gg+gq_{S}$ and all scattering
channels. The upper frame shows cross sections for the default choice
of scales specified in Section~\protect\ref{subsection:CompleteResummed}
(solid), as well as for varied scales $C_{2}=C_{4}=2$ (dashes), $C_{2}=C_{4}=0.5$
(dots), and $C_{3}=4e^{-2\gamma_{E}}$ (dot-dashes). The lower two
frames show ratios of the cross sections computed for the varied factorization
scales to the cross section for the default choice of the scales. }

\label{Fig:ScaleTev2} 
\end{figure*}

To illustrate the relative importance of the individual initial-state
contributions in the final answer, we provide a parton flavor decomposition
of our resummed transverse momentum distribution $d\sigma/dQ_{T}$
in Fig.~\ref{Fig:bStarCDF}. This distribution is integrated over
all diphoton invariant masses $Q$, subject to the CDF cuts, and receives
dominant contributions from the $Q_{T}<Q$ region. The $gg+gq_{S}$
contribution supplies about one-third of the total rate near $Q_{T}=5$~GeV.
It falls steeply after $Q_{T}>20$ GeV, because the gluon PDF falls
steeply with parton fractional momentum $x$.

Dependence of the resummed cross sections on the choice of factorization
scales mentioned in Section~\ref{subsection:CompleteResummed} is
examined in Fig.~\ref{Fig:ScaleTev2}. We pick a few characteristic
combinations of alternative scales to probe the scale dependence associated
with the resummed Sudakov function $e^{-{\cal S}}$, the $b$-dependent
PDF's ${\mathcal{P}}_{a/h}(x,\vec{b})\approx[{\mathcal{C}}_{a/c}\otimes f_{c/h}](x,b;\mu)$,
and the regular $Y$ term. The small-$Q_{T}$ region is sensitive
primarily to the scales $C_{1}/b,C_{2}\, Q,C_{3}/b$ in the resummed
term $W$. The event rate at large $Q_{T}$ is controlled by the choice
of the factorization scale $\mu_{F}\equiv C_{4}\, Q$ in the regular
term $Y$.

At the relatively low values of $Q$ relevant for the Tevatron experiments,
the scale dependence of the next-to-leading order $gg+gq_{S}$ cross
section is still substantial, with variations being about $-20\%$
($+50\%$) at $Q_{T}=5-10$ GeV, $\pm10\%$ at $Q_{T}=10-20$ GeV,
and $\pm20\%$ at $Q_{T}=20-40$ GeV. Since the $Y$ term is the lowest-order
approximation for $gg\rightarrow\gamma\gamma g$ at $Q_{T}\sim Q$,
the scale dependence associated with the constant $C_{4}$ remains
pronounced at large $Q_{T}$. The inclusive $gg+gq_{S}$ rate, integrated
over $Q_{T}$, varies by $20-40\%$ almost independently of the $\gamma\gamma$
invariant mass $Q$. The large scale dependence of the NLO $gg+gq_{S}$
cross section reflects slow perturbative convergence in gluon gluon
scattering, observed also in other similar processes, e.g., $gg\rightarrow\mbox{Higgs}$
via the top quark loop \cite{Harlander:2002wh,Anastasiou:2004xq,Anastasiou:2005qj}.
For this reason, a NNLO calculation would be desirable to reduce the
scale uncertainty in the $gg+gq_{S}$ channel.

On the other hand, the scale dependence of the cross section when
all channels are combined is relatively mild, with variations not
exceeding 10\% at small $Q_{T}$ and 20\% at large $Q_{T}$. Variations
in the integrated inclusive rate for all channels combined are below
10\% at $Q>30$ GeV.

\begin{figure*}
\includegraphics[width=1\textwidth,height=6.5cm,keepaspectratio]{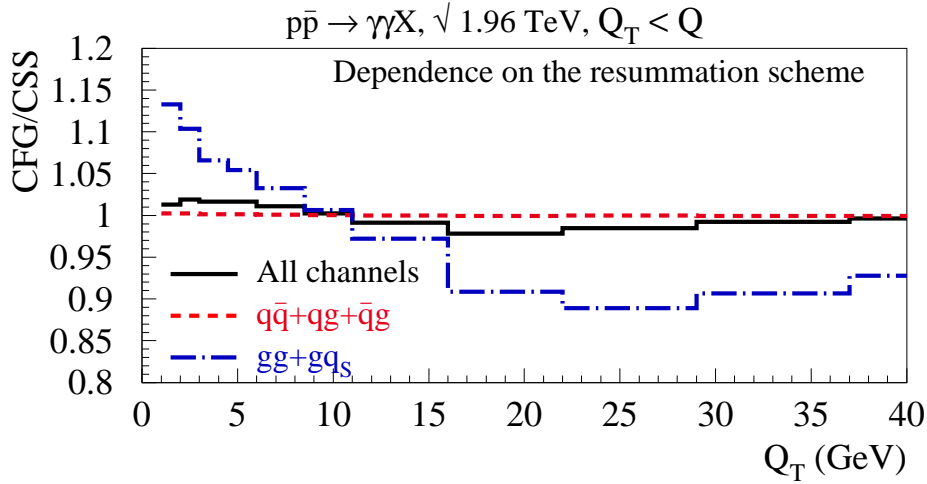}

(a)

\includegraphics[width=1\textwidth,height=6.5cm,keepaspectratio]{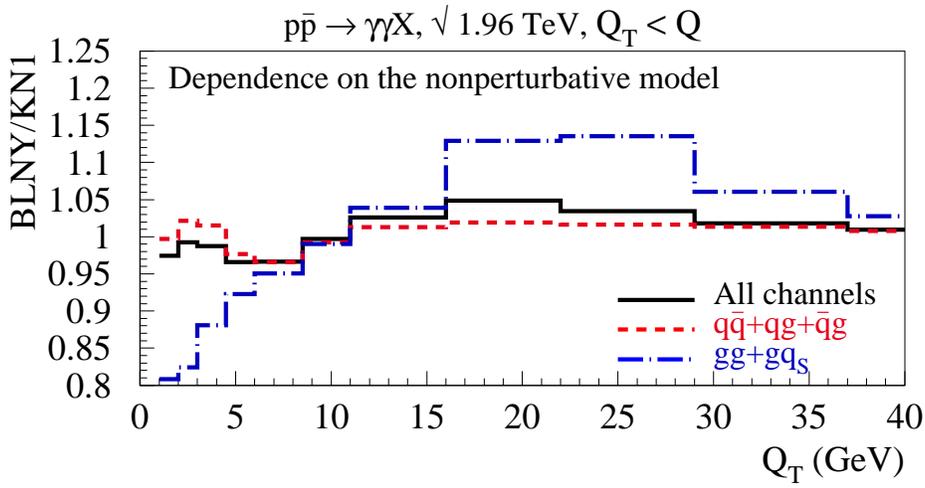}

(b)

\caption{Ratios of resummed cross sections at the Tevatron Run-2 computed
in (a) the Catani-de Florian-Grazzini (CFG) and Collins-Soper-Sterman
(CSS) resummation schemes and (b) using the BLNY and KN1 nonperturbative
models, as functions of the $\gamma\gamma$ transverse momentum $Q_{T}$.
The ratios are shown in the $q\bar{q}+qg$ (dashed), $gg+gq_{S}$
(dot-dashed), and all (solid) scattering channels. A $Q_{T}<Q$ cut
is imposed in this comparison.}

\label{Fig:VariationsTev2} 
\end{figure*}

Another aspect of scale dependence is associated with the assumed
arrangement of logarithmic terms in the resummed $W$ term, i.e.,
the {}``resummation scheme'' that is adopted. This dependence is
yet another indicator of the size of higher-order corrections not
included in the present analysis. Figure~\ref{Fig:VariationsTev2}(a)
shows ratios of the full resummed cross sections in the Catani-de
Florian-Grazzini (CFG) and Collins-Soper-Sterman (CSS) resummation
schemes, as described in Sec.~III. The differences between these
schemes stem from the different treatment of the NLO hard-vertex correction
$h_{a}^{(1)}(\theta_{*})$. The magnitude of $h_{a}^{(1)}(\theta_{*})$
determines whether the channel is sensitive to the choice of the two
resummation schemes. The magnitude of $h_{g}^{(1)}(\theta_{*})$ in
the $gg+gq_{S}$ channel exceeds that of $h_{q}^{(1)}(\theta_{*})$
in the $q\bar{q}+qg$ channel by roughly an order of magnitude for
most values of the $\theta_{*}$ angle \cite{Nadolsky:2002gj}. Consequently,
while the dependence on the resummation scheme is practically negligible
in the dominant $q\bar{q}+qg$ channel (dashed line), it can reach
15\% in the subleading $gg+gq_{S}$ channel (dot-dashed line). The
$Q_{T}$ spectrum in $gg+gq_{S}$ channel is slightly softer in the
CFG scheme up to the point of switching to the fixed-order cross section
at $Q_{T}\approx60$ GeV. The resummation scheme dependence in all
channels (solid line) is less than 3-4\%, reflecting mostly the scheme
dependence in the $gg+gq_{S}$ channel.

To examine the sensitivity of the resummed predictions to long-distance
nonperturbative dynamics in hadron-hadron scattering, we include in
Fig.~\ref{Fig:VariationsTev2}(b) a comparison with the resummed
cross sections for an alternative choice of the nonperturbative model.
As explained in Sec.~\ref{subsection:CompleteResummed}, our default
calculation is performed in the recent KN1 model \cite{Konychev:2005iy}
for the nonperturbative part of the resummed form factor $\widetilde{W}(Q,b,y,\Omega_{*})$.
Figure~\ref{Fig:VariationsTev2}(b) shows ratios of the predictions
for a different BLNY model \cite{Landry:2002ix} and our default KN1
model in various initial-state scattering channels.

The difference is maximal at the lowest $Q_{T}$, as expected, and
it is less than 5\% for the total cross section. For the $q{\bar{q}}+qg$
and $gg+gq_{S}$ initial states the maximal difference is about 5\%
and 20\%, respectively. The dependence on the nonperturbative function
is stronger in the $gg+gq_{S}$ channel, where the BLNY/KN1 ratio
in the $gg+gq_{S}$ channel reaches its maximum of 1.15 at $Q_{T}\approx25$
GeV and slowly decreases toward 1, reached at the switching point
at $Q_{T}\approx60$ GeV. This behavior reflects our assumption of
a larger magnitude of the nonperturbative function in the $gg+gq_{S}$
channel, which is rescaled in our model by $C_{A}/C_{F}=9/4$ compared
to the nonperturbative function in the $q\bar{q}+qg$ channel. In
summary, despite a few-percent uncertainty associated with the nonperturbative
function in the $gg+gq_{S}$ process, the overall dependence of the
Tevatron $\gamma\gamma$ cross section on the nonperturbative input
can be neglected.

\subsection{Results for the LHC}

To obtain predictions for $pp$ collisions at the LHC at $\sqrt{S}=14$~TeV,
we employ the cuts on the individual photons used by the ATLAS collaboration
in their simulations of Higgs boson decay, $h\rightarrow\gamma\gamma$~\cite{ATLAS:1999fr}.
We require transverse momentum~$p_{T}^{\gamma}>40~(25)$~GeV for
the harder (softer) photon, and rapidity~$|y^{\gamma}|<2.5$ for
each photon. We impose the ATLAS isolation criteria, looser than for
the Tevatron study, requiring less than 15~GeV of hadronic and extra
electromagnetic transverse energy inside a $\Delta R=0.4$ cone around
each photon. We also require the separation $\Delta R_{\gamma\gamma}$
between the two isolated photons to be above 0.4. The cuts optimized
for the Higgs boson search may require adjustments in order to test
perturbative QCD predictions in the full $\gamma\gamma$ invariant
mass range accessible at the LHC.

\begin{figure*}
\includegraphics[width=0.49\textwidth]{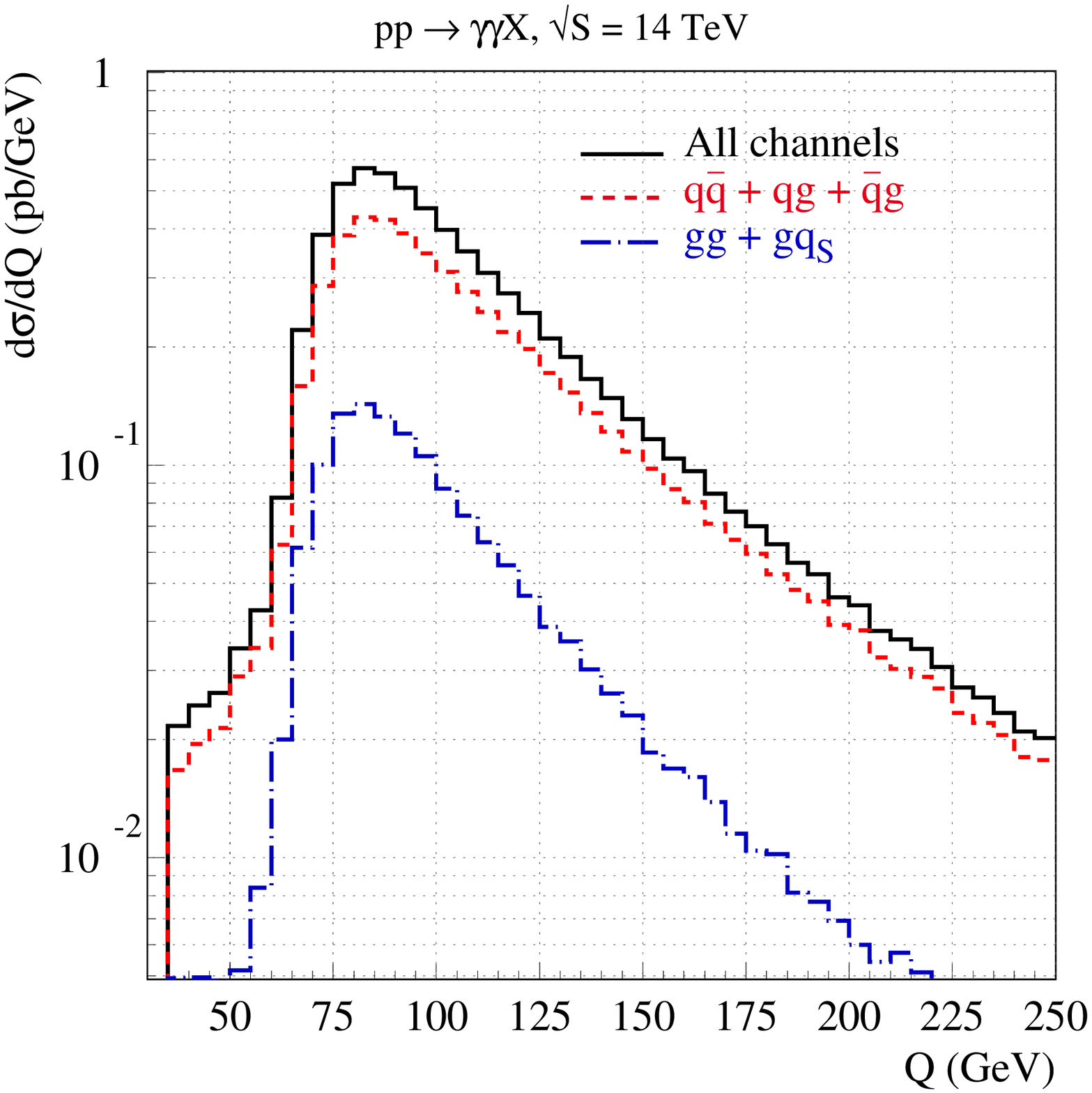}\includegraphics[width=0.49\textwidth]{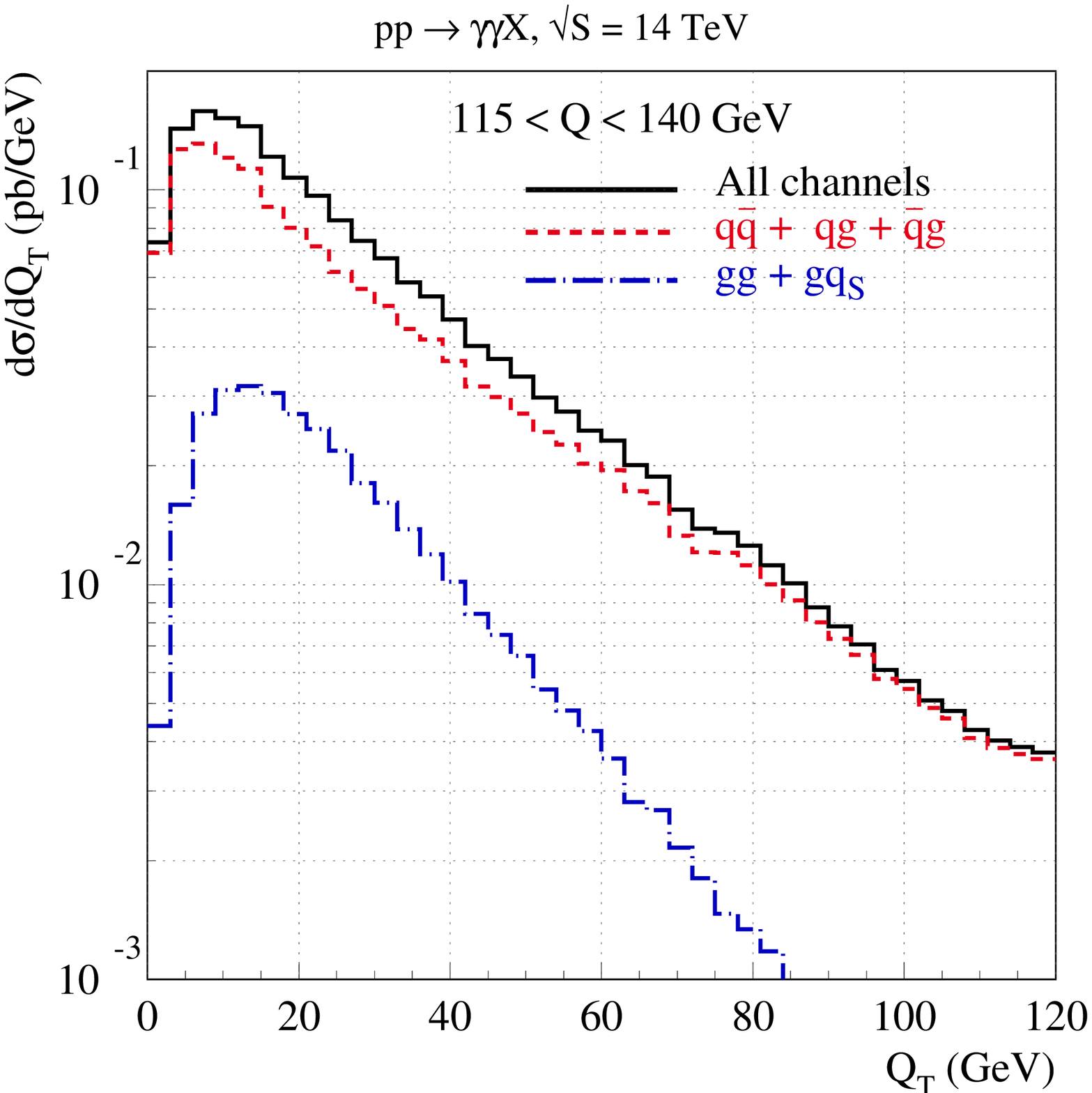}\\
(a)\hspace{3in}(b)

\includegraphics[width=0.49\textwidth]{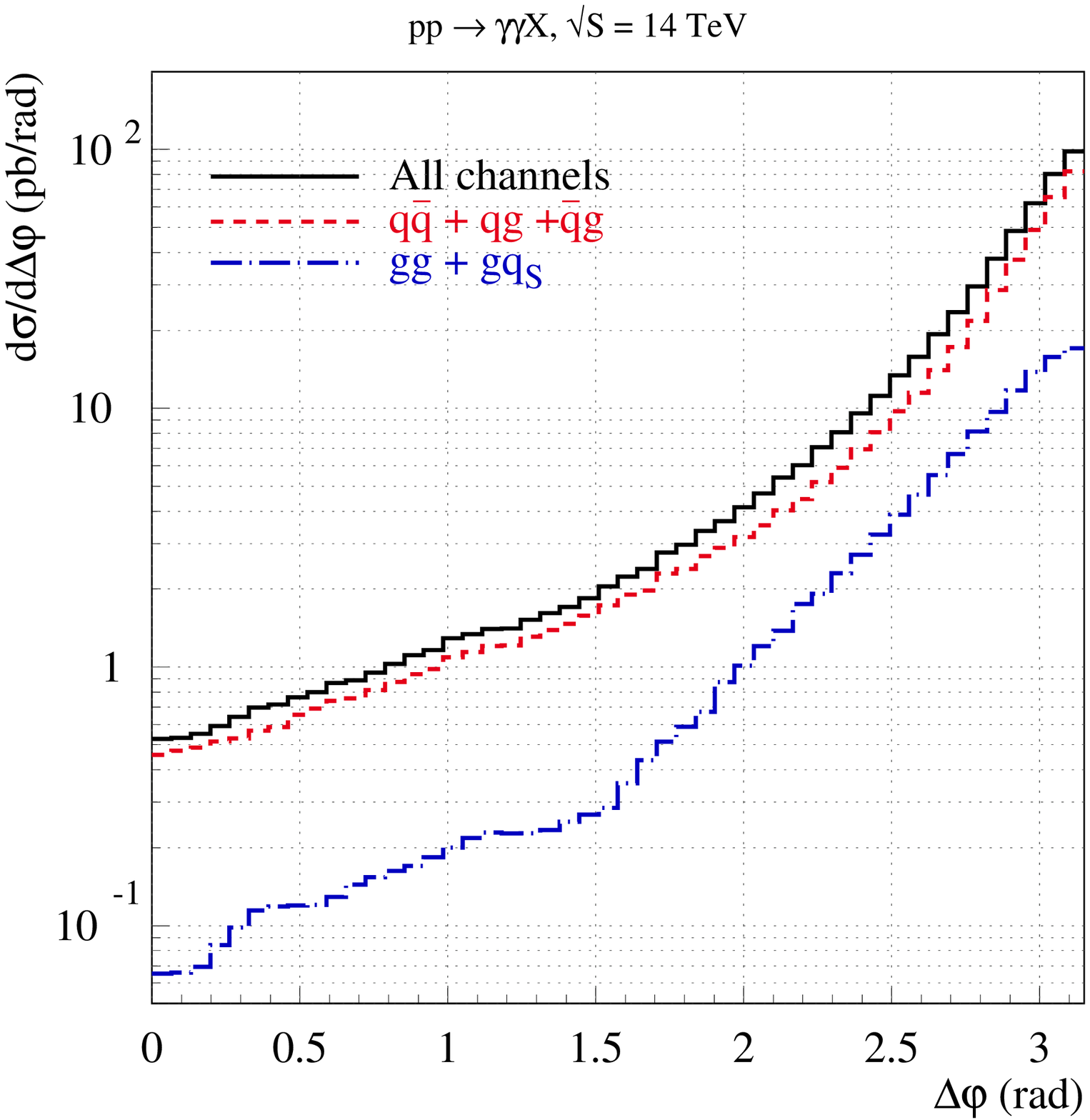}\\
(c)

\caption{Resummed $d\sigma/dQ,$ $d\sigma/dQ_{T},$ and $d\sigma/d\Delta\varphi$
distributions of photon pairs at the LHC for ATLAS kinematic cuts.
\label{Fig:QDelPhiLHC}}
\end{figure*}

Distributions in the invariant mass $Q$, transverse momentum $Q_{T}$,
and azimuthal angle separation $\Delta\varphi\equiv\varphi_{\gamma_{1}}-\varphi_{\gamma_{2}}$
between the two photons in the laboratory frame are shown in Fig.~\ref{Fig:QDelPhiLHC}.
As before, we compare the magnitudes of the $q\bar{q}+qg$ and $gg+gq_{S}$
cross sections. The qualitative features are similar to those at the
Tevatron, but the relative contribution of the various initial states
changes at the LHC. The $gg+gq_{S}$ initial state contributes about
25\% of the total rate at $Q\sim80$~GeV where the mass distribution
peaks, but the $gg+gq_{S}$ rate falls faster than $q\bar{q}+qg$
with increasing invariant mass.

In the invariant mass range relevant for the Higgs boson search, $115<Q<140$
GeV, the transverse momentum distribution in Fig.~\ref{Fig:QDelPhiLHC}(b)
shows that the $gg+gq_{S}$ initial state accounts for about 25\%
of the rate at low $Q_{T}$. At high transverse momentum, on the other
hand, the other channels dominate. The relative size of the $gg+gq_{S}$
contribution drops as the invariant mass or the transverse momentum
of the photon pair grows. The $gg+gq_{S}$ contribution falls more
steeply with $Q_{T}$ for larger masses of the diphoton. These features
are attributable to the steeply falling gluon distribution as a function
of increasing momentum fraction $x$.

\begin{figure*}
\includegraphics[width=0.7\textwidth,keepaspectratio]{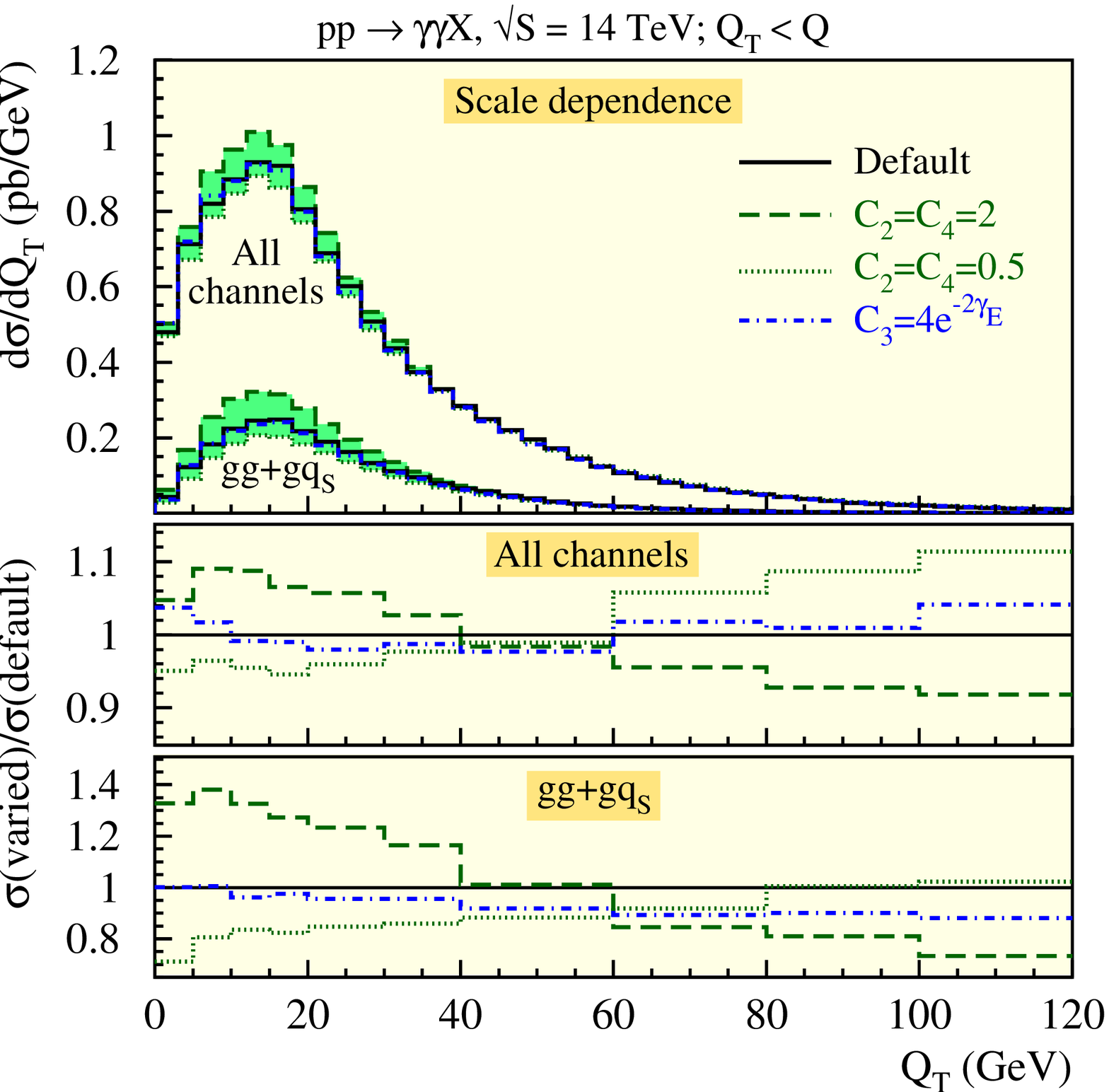}

\caption{Scale dependence in the $gg+gq_{S}$ and all scattering channels
at the LHC for the same scale choices as in Fig.~\protect\ref{Fig:ScaleTev2}.}

\label{Fig:ScaleLHC} 
\end{figure*}

The scale dependence at the LHC, presented in Fig.~\ref{Fig:ScaleLHC},
is somewhat reduced compared to the Tevatron (cf. Fig.~\ref{Fig:ScaleTev2}).
Maximum scale variations of about 40\% in the $gg+gq_{S}$ channel
are observed at the peak of the $d\sigma/dQ_{T}$ distribution, and
they are substantially smaller at large $Q_{T}$. The scale variation
in the sum over all channels does not exceed 10\% (15\%) at small
$Q_{T}$ (large $Q_{T}$). Variations in the integrated inclusive
rate at $Q>50$ GeV are below 7\% (30\%) in all channels ($gg+gq_{S}$
channel).

\begin{figure*}
\includegraphics[width=1\textwidth,height=6.5cm,keepaspectratio]{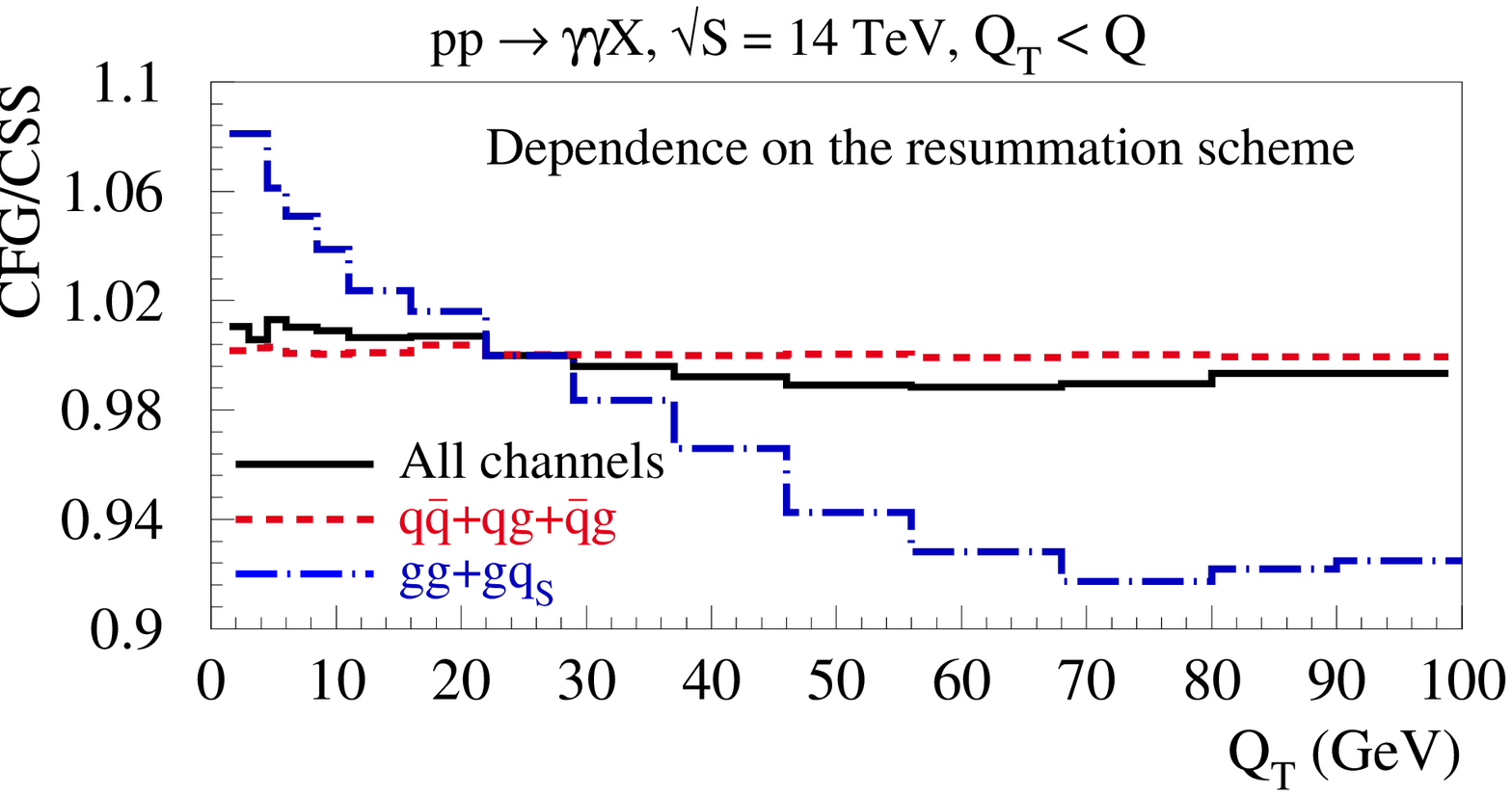}\\
(a)

\includegraphics[width=1\textwidth,height=6.5cm,keepaspectratio]{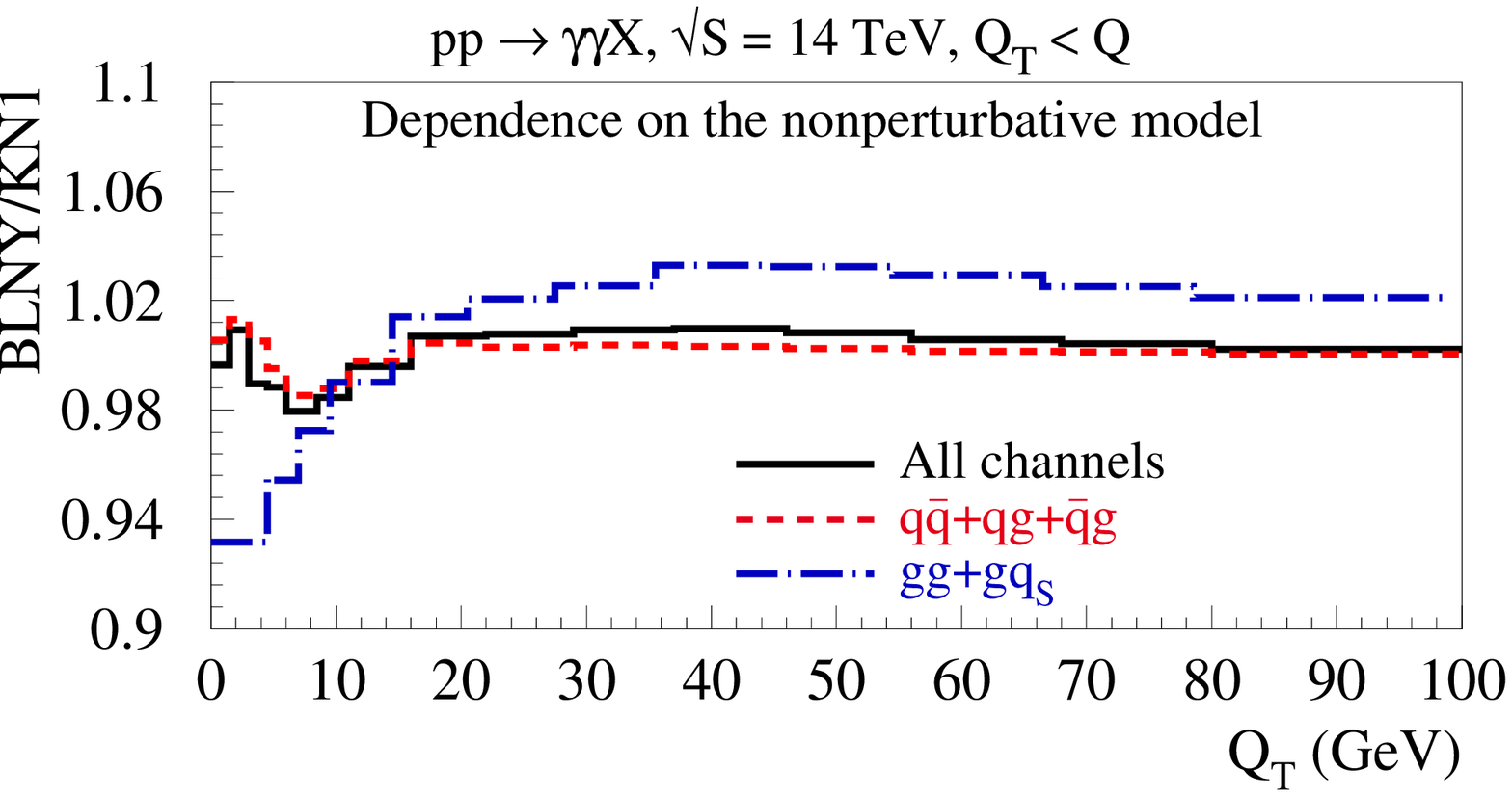}\\
(b)

\caption{Same as Fig.~\ref{Fig:VariationsTev2}, at the LHC. \label{Fig:VariationsLHC}}
\end{figure*}

The dependence on the resummation scheme is mild at the LHC (cf.~Fig.~\ref{Fig:VariationsLHC}(a)),
with the maximal differences between the CSS and CFG schemes below
0.5\%, 10\%, and 2\% in $q\bar{q}+qg$, $gg+gq_{S}$, and all channels.
The scheme dependence is again the largest in the $gg+gq_{S}$ channel,
where it persists up to the point of switching to the fixed-order
cross section at $Q_{T}\approx120$ GeV. The ratios of the resummed
cross sections calculated in the BLNY and KN1 models for nonperturbative
contributions in the CSS scheme are shown in Fig.~\ref{Fig:VariationsLHC}(b).
The influence of the long-distance (large-$b)$ contributions is suppressed
at the high center-of-mass energy of the LHC. Differences between
the predictions in the two models do not exceed 2\%, 6\%, and 2\%
in the $q\bar{q}+qg,$ $gg+gq_{S},$ and all scattering channels.

The KN1 and BLNY nonperturbative models neglect the possibility of
a strong $x$ dependence of the nonperturbative function, which may
substantially modify our predictions at the energy of the LHC collider.
Analysis of small-$x$ semi-inclusive deep inelastic scattering data
\cite{Berge:2004nt} suggests that $x$-dependent nonperturbative
corrections of uncertain magnitude may substantially affect the resummed
cross sections. Such corrections can be constrained by studying the
rapidity and energy dependence of the nonperturbative function at
the Tevatron and LHC, for example, from copious production of $Z$
bosons \cite{Berge:2004nt}. We conclude that uncertainties due to
the choice of the resummation scheme and the nonperturbative model
will be small at the LHC, if the resummed nonperturbative function
does not vary strongly with $x$.

\begin{figure}
\includegraphics[width=0.5\textwidth,keepaspectratio]{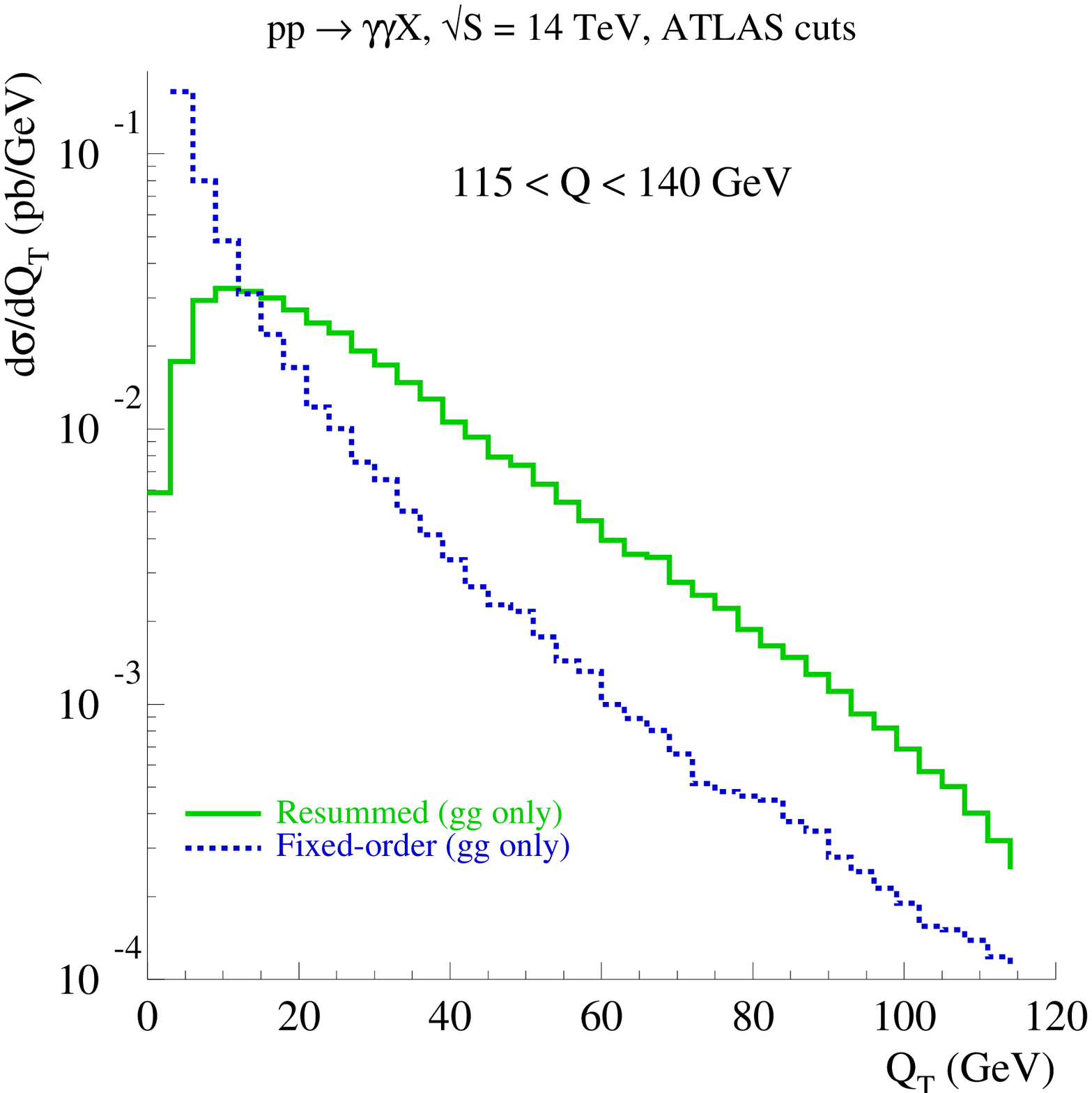}\includegraphics[width=0.5\textwidth,keepaspectratio]{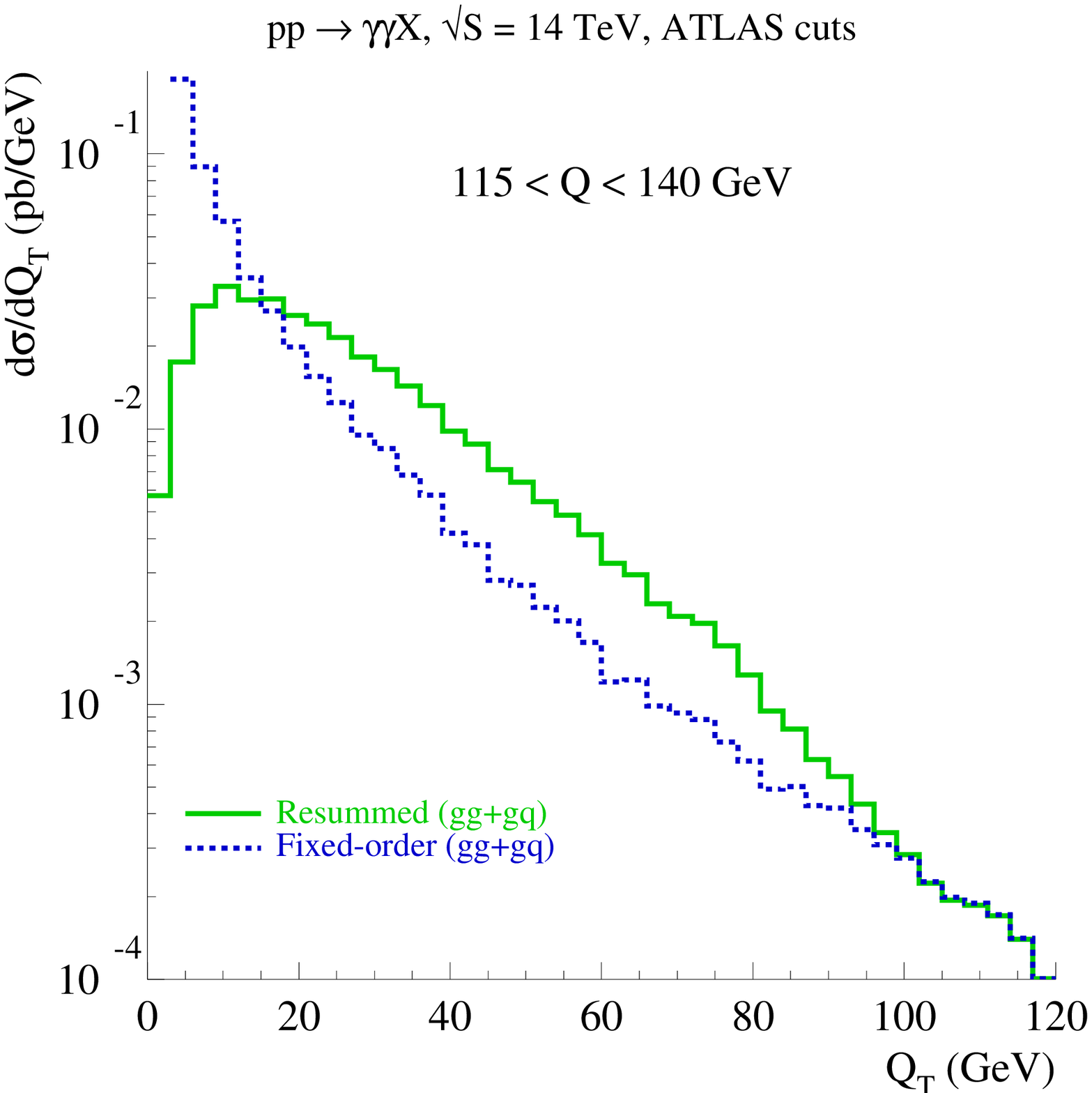}\\
(a)\hspace{3in}(b)

\caption{Inclusion of the $qg$ contribution improves the matching of the
resummed and NLO perturbative cross sections at large $Q_{T}$, as
demonstrated by these plots of the resummed and finite-order NLO cross
sections for (a) the $gg$ channel only; (b) the combined $gg+gq_{S}$
channel. The resummed and NLO cross sections are shown by the solid
and dashed lines. \label{Fig:gqSLHC}}
\end{figure}

\subsection{The role of the $gq_{S}$ contribution \label{subsection:gqS}}

Figures~\ref{Fig:bStarCDF}-\ref{Fig:VariationsLHC} show the contributions
from the $q\bar{q}+qg$ and $gg+gq_{S}$ channels along with their
sum. One may wonder if a further decomposition into $q\bar{q}$ and
$qg$ (or $gg$ and $gq_{S}$) contributions could provide additional
insights into the relative importance of different scattering processes.
We observe in our calculations that the resummed cross sections $W+Y$
and the fixed-order cross sections $P$ in the elementary scattering
subchannels ($q\bar{q}$, $qg$,...) may not cross until $Q_{T}$
is significantly larger than $Q$. This result is at variance with
our expectation that the fixed-order answer should be adequate when
$Q_{T}$ is of order $Q$, where logarithmic effects are small, and
the one-scale nature of the dynamics seems apparent.

Consider, for example, the $gg$ and $gg+gq_{S}$ transverse momentum
distributions in the mass interval $115<Q<140$ GeV at the LHC shown
in Figs.~\ref{Fig:gqSLHC}(a) and (b). In the $gg$ channel alone
(Fig.~\ref{Fig:gqSLHC}(a)), the $W+Y$ cross section remains above
the NLO cross section $P$ until $Q_{T}\sim140\mbox{ GeV}$. However,
after the $gq_{S}$ contribution is included (Fig.~\ref{Fig:gqSLHC}(b)),
$W+Y$ crosses $P$ at $Q_{T}\sim105\mbox{ GeV}$. Our expectation
of the adequacy of the NLO prediction at $Q_{T}\sim Q$ is satisfied
in this case, and this conclusion also holds for other intervals of
$Q$. At the crossing point, the two cross sections satisfy $W+Y=P,$
i.e., $\textrm{$W=A$}$; the resummed term is equal to its NLO perturbative
expansion, the asymptotic term. Similarly, good matching of the resummed
and NLO cross sections in the $q\bar{q}+qg$ channel requires that
we include both $q\bar{q}$ and $qg$ contributions.

This feature can be understood by noticing that the flavors of the
PDF's $f_{a/h}(x,\mu)$ mix in the process of PDF evolution. Consequently
the perturbative expansion of $W$ in the $gg$ channel contains the
full NLO asymptotic piece $A$ in the combined $gg+gq_{S}$ channel,
generated from the Sudakov exponential and lowest-order resummed contribution
$\propto f_{g/h_{1}}(x_{1},1/b)\, f_{g/h_{2}}(x_{2},1/b)$ evaluated
at a scale of order $1/b$. The mismatch between the flavor content
in the perturbatively expanded $W$ and $A$ in nominally the same
$gg$ subchannel causes the difference $W-A$ to be large and delays
the crossing. On the other hand, the flavor content of $W$ and $A$
is the same (up to NNLO) when the $gg$ and $gq_{S}$ contributions
are combined, and the matching is improved. The $gq_{S}$ scattering
subchannel has been assumed to be small and neglected in past studies,
and indeed it contributes about one tenth of the $gg+gq_{S}$ inclusive
rate $d\sigma/dQ$. However, we see that the $gq_{S}$ contribution
must be included to correctly predict $d\sigma/dQ_{T}$ and to realize
matching between the resummed and perturbative contributions at large
transverse momenta.

\section{Summary and Conclusions}

In this paper, we address new theoretical issues in $Q_{T}$ resummation
at two-loop accuracy that arise in the gluon-gluon subprocess, $gg+gq\rightarrow\gamma\gamma$,
one of the important short-distance subprocesses that contribute to
the inclusive reactions $p\bar{p}\rightarrow\gamma\gamma X$ at the
Fermilab Tevatron and $pp\rightarrow\gamma\gamma X$ at the CERN Large
Hadron Collider (LHC).

We evaluate all next-to-leading (NLO) contributions of order ${\mathcal{O}}(\alpha^{2}\alpha_{s}^{3})$
to the $gg+gq\rightarrow\gamma\gamma$ process (Fig.~\ref{Fig:FeynDiag}(b-e)).
A new ingredient in this paper is the inclusion of the $gq\rightarrow\gamma\gamma q$
process, Fig.~\ref{Fig:FeynDiag}(d), a necessary component of the
resummed NLO contribution. We resum to next-to-next-to-leading logarithmic
(NNLL) accuracy the large logarithmic terms of the form $\ln(Q_{T}^{2}/Q^{2})$
in the limit when $Q_{T}$ of the $\gamma\gamma$ pair is smaller
than its invariant mass $Q$. The perturbative Sudakov functions ${\mathcal{A}}$
and ${\mathcal{B}}$ and the Wilson coefficient functions ${\mathcal{C}}$
in the resummed cross section $W$ are computed to orders $\alpha_{s}^{3},$
$\alpha_{s}^{2}$, and $\alpha_{s}$. The resummed cross sections
are computed according to the CSS \cite{Collins:1984kg} and CFG \cite{Catani:2000vq}
resummation schemes, with the differences between the two approaches
reflecting the size of higher-order corrections. A new nonperturbative
function \cite{Konychev:2005iy}, dominated by a process-independent
soft correction, is employed to describe the dynamics at large impact
parameters.

Subtraction of the singular logarithmic contributions associated with
initial-state radiation from the NLO cross section $P$ defines a
regular piece $Y$. This regular term is added to the small-$Q_{T}$
resummed cross section $W$ to predict the production rate at small
to moderate values of $Q_{T}$. In the $gg$ channel, we also subtract
from $P$ a new singular spin-flip contribution that affects azimuthal
angle ($\varphi_{*})$ dependence in the Collins-Soper reference frame.
For our final prediction, we switch from the resummed cross section
$W+Y$ to $P$ at the point where $W+Y$ crosses $P$, approaching
$P$ from above, as in Ref.~\cite{Balazs:1997xd}. The location of
this point in $Q_{T}$ is of order $Q$ in the $q\bar{q}+qg$ and
the $gg+gq$ channels. For such matching to happen, it is essential
to combine cross sections in the $q\bar{q}$ and $qg$ ($gg$ and
$gq$) channels, as demonstrated in Sec.~\ref{subsection:gqS}.

At the LHC (Tevatron), the $gg+gq$ subprocess contributes 20\% (10\%)
of the total $\gamma\gamma$ production rate (integrated over the
full range of the photons' momenta). The relative contribution of
$gg+gq$ scattering may reach 25\% for some $Q$ and $Q_{T}$ values.
The $gg+gq$ channel provides an interesting opportunity to test CSS
resummation at a loop level and may be explored in detail at later
stages of the LHC operation. The NNLL/NLO resummed cross section for
the $gg+gq_{S}$ channel is used in Ref.~\cite{Balazs:2007hr} to
predict fully differential distributions of Higgs bosons and QCD background
at the LHC in the $\mbox{Higgs}\rightarrow\gamma\gamma$ decay mode.

\section*{Acknowledgments}

Research in the High Energy Physics Division at Argonne is supported
in part by US Department of Energy, Division of High Energy Physics,
Contract DE-AC02-06CH11357. The work of C.-P. Y. is supported by the
U. S. National Science Foundation under grant PHY-0555545. P.M.N.
thanks G. Bodwin for discussions of the azimuthal angle dependence
in gluon scattering. We gratefully acknowledge the use of \textit{Jazz},
a 350-node computing cluster operated by the Mathematics and Computer
Science Division at ANL as part of its Laboratory Computing Resource
Center. The Feynman diagrams in Fig.~\ref{Fig:FeynDiag} were drawn
with aid of the program \textsc{JaxoDraw} \cite{Binosi:2003yf}.

\appendix

\section{The $gq_{S}\rightarrow\gamma\gamma q_{S}$ amplitude\label{Appendix:qgAAq}}

To obtain the gluon-quark contribution to the $gg+gq_{S}$ scattering
channel shown in Fig.~\ref{Fig:FeynDiag}(d), we derive the helicity-dependent
$q\bar{q}\gamma\gamma g\rightarrow0$ amplitude ${\mathcal{M}}_{5}(q_{1},\bar{q}_{2},\gamma_{3},\gamma_{4},g_{5})$
from the one-loop $q\bar{q}ggg$ amplitude in the color-decomposed
representation available in Ref.~\cite{Bern:1994fz}. The $q\bar{q}\gamma\gamma g$
amplitude is expressed as\begin{equation}
{\mathcal{M}}_{5}(q_{1}^{c_{1}},\bar{q}_{2}^{c_{2}},\gamma_{3},\gamma_{4},g_{5}^{a_{5}})=2g^{3}e^{2}\left(\sum_{i_{l}}e_{i_{l}}^{2}\right)T_{c_{1}c_{2}}^{a_{5}}\sum_{\sigma\in S_{3}^{(345)}}A_{5;1}^{L,[1/2]}(1_{q},2_{\bar{q}};\sigma(3),\sigma(4),\sigma(5))\label{A5qqAAg}\end{equation}
 in terms of the primitive amplitudes $A_{5;1}^{L,[1/2]}(1_{q},2_{\bar{q}};3,4,5)=-A_{5;1}^{f}(1_{q},2_{\bar{q}};3,4,5)-A_{5;1}^{s}(1_{q},2_{\bar{q}};3,4,5)$
for $q\bar{q}ggg\rightarrow0$ scattering involving a spin-1/2 fermion
loop. The amplitude ${\mathcal{M}}_{5}$ is proportional to the sum
$\sum_{i_{l}}\left(e^{2}e_{i_{l}}^{2}\right)$ of squared quark charges
circulating in the fermion loop, as well as the QCD generator matrix
$T_{c_{1}c_{2}}^{a_{5}},$ with $\mbox{{Tr}}(T^{a_{1}}T^{a_{2}})=\delta^{a_{1}a_{2}}$.
The color indices $c_{1},\, c_{2},$ and $a_{5}$ belong to the quark
1, antiquark 2, and gluon 5. The primitive amplitudes are summed over
all possible permutations $S_{3}^{(345)}$ of the legs 3, 4, and 5.

Equation~(\ref{A5qqAAg}) is derived from Eq.~(2.10) of Ref.~\cite{Bern:1994fz}
after gluons 3 and 4 are replaced with photons, i.e., the QCD generators
$T^{a_{3}}$and $T^{a_{4}}$ are replaced by identity matrices, and
the overall charge factor is adjusted, $g^{5}\rightarrow2g^{3}\sum_{i_{l}}(ee_{i_{l}})^{2}.$
It correctly reproduces the small-$Q_{T}$ asymptotic behavior reflected
in Eq.~(\ref{ASYgg}), which we derive by applying factorization
relations in the splitting amplitude formalism discussed in Appendix~\ref{Appendix:ASYgg}.
The $q\bar{q}\gamma\gamma g$ amplitude in Eq.~(\ref{A5qqAAg}) disagrees
with the one published in Ref.~\cite{Yasui:2002bn} which appears
to violate factorization relations in the $q\parallel\bar{q}$ limit.
A few independent amplitudes $A_{5;1}^{f}(1_{q},2_{\bar{q}};3,4,5)$
and $A_{5;1}^{s}(1_{q},2_{\bar{q}};3,4,5)$ are presented explicitly
in Sec.~5 of Ref.~\cite{Bern:1994fz}, with the remaining amplitudes
related by discrete symmetries according to Eq.~(5.25) in that publication.
Some $q\bar{q}ggg$ amplitudes contain infrared poles, which cancel
in the sum over permutations $S_{3}^{(345)}$. We retain only non-vanishing
finite parts $F^{x}$ of such divergent amplitudes, i.e., we take
$A_{5;1}^{x}=iF^{x}/(16\pi^{2})$ for $x=f$ and $s.$

\section{Derivation of the small-$Q_{T}$ asymptotic term for gluon-gluon
scattering \label{Appendix:ASYgg}}

In this appendix, we derive the small-$Q_{T}$ asymptotic approximation
Eq.~(\ref{ASYgg}) for the NLO cross section in $g_{1}g_{2}\rightarrow\gamma_{3}\gamma_{4}$
scattering. We expand the finite-order cross section as a series in
the small parameter $Q_{T}^{2}/Q^{2}$. Consider first the leading
real-emission contributions, which arise when gluon 5 is radiated
off the external gluon leg 1 or 2 as in Fig.~\ref{Fig:FeynDiag}(b).%
\footnote{In contrast, Feynman graphs with gluon radiation off a propagator
in the quark loop {[}Fig.~\ref{Fig:FeynDiag}(c)] are finite in the
$Q_{T}\rightarrow0$ limit.%
} In the notation introduced in Sec.~III, the small-$Q_{T}$ approximation
for the real-emission cross section takes the form\begin{eqnarray}
 &  & \left.A(Q,Q_{T},y,\Omega_{*})\right|_{real}=\int_{x_{1}}^{1}d\xi_{1}\int_{x_{2}}^{1}d\xi_{2}f_{g/h_{1}}(\xi_{1},\mu_{F})f_{g/h_{2}}(\xi_{2},\mu_{F})\frac{1}{(2\pi)^{4}}\frac{1}{64\xi_{1}\xi_{2}S}\left|{\mathcal{M}}_{5}\right|^{2}\nonumber \\
 & \times & \left\{ \frac{\delta(\xi_{1}-x_{1})}{\left[1-\widehat{x}_{2}\right]_{+}}+\frac{\delta(\xi_{2}-x_{2})}{\left[1-\widehat{x}_{1}\right]_{+}}-x_{1}x_{2}\delta(\xi_{1}-x_{1})\delta(\xi_{2}-x_{2})\ln\frac{Q_{T}^{2}}{Q^{2}}\right\} .\label{Asygg1}\end{eqnarray}
 The right-hand side of Eq.~(\ref{Asygg1}) includes a product of
the gluon parton densities $f_{g/h_{i}}(\xi_{1,2},\mu_{F}),$ squares
of parton-scattering amplitudes ${\mathcal{M}}_{5}$, and phase-space
factors, integrated over the light-cone momentum fractions $\xi_{1,2}\equiv p_{1,2}^{+}/P_{1,2}^{+}$
of the incoming gluons 1 and 2. The delta-functions constrain integration
to phase-space regions where the final-state gluon 5 is collinear
to gluon 1 {[}$p_{5}^{\mu}=(1-\widehat{x}_{1})p_{1}^{\mu}$], collinear
to gluon 2 {[}$p_{5}^{\mu}=(1-\widehat{x}_{2})p_{2}^{\mu}$], or soft
{[}$p_{5}^{\mu}\rightarrow0$], with $\widehat{x}_{i}\equiv x_{i}/\xi_{i}$
for $i=1,2$.

The $2\rightarrow3$ helicity amplitude ${\mathcal{M}}_{5}(1,2,3,4,5)$
is analyzed conveniently in an unphysical scattering channel $0\rightarrow g(\bar{p}_{1},\bar{\lambda}_{1})g(\bar{p}_{2},\bar{\lambda}_{2})\gamma(\bar{p}_{3},\bar{\lambda}_{3})\gamma(\bar{p}_{4},\bar{\lambda}_{4})g(\bar{p}_{5},\bar{\lambda}_{5})$.
The momenta $\bar{p}_{i}$ and helicities $\bar{\lambda}_{i}$ are
related to the physical momenta $p_{i}$ and helicities $\lambda_{i}$
as $\{\barp_{i},\barl_{i}\}$=$\{-p_{i},-\lambda_{i}\}$ for $i=1$
or 2, and $\{\barp_{i},\barl_{i}\}$=$\{ p_{i},\lambda_{i}\}$ for
$i=$3, 4, or 5. ${\mathcal{M}}_{5}(1,2,3,4,5)$ is a shorthand notation
for ${\mathcal{M}}_{5}(\barp_{1},\barl_{1};\barp_{2},\barl_{2};\barp_{3},\barl_{3};\barp_{4},\barl_{4};\barp_{5},\barl_{5}).$

The amplitude ${\mathcal{M}}_{5}(1,2,3,4,5)$ was derived in Refs.~\cite{Balazs:1999yf,deFlorian:1999tp}
from color-decomposed 5-gluon 1-loop scattering amplitudes \cite{Bern:1993mq}.
${\mathcal{M}}_{5}(1,2,3,4,5)$ is built from 1-loop partial amplitudes
$A_{5;1}(1,2,3,4,5)$ for the $0\rightarrow gg\gamma\gamma g$ scattering
process, identical to the partial amplitudes for $0\rightarrow ggggg$
scattering via a spin-1/2 fermion loop \cite{Bern:1993mq}. The squared
5-leg amplitude, averaged over spins, colors, and identical final-state
particles, is\begin{eqnarray}
|{\mathcal{M}}_{5}|^{2} & = & \sigma_{g}^{(1)}\sum_{\barl_{1},\bar{\lambda}_{2},\bar{\lambda}_{3},\bar{\lambda}_{4},\bar{\lambda}_{5}}\Biggl|\,\sum_{\sigma\in{\textrm{COP}}_{3}^{(125)}}A_{5;1}(\sigma_{1},\sigma_{2},\sigma_{3},\sigma_{4},\sigma_{5})\Biggr|^{2},\label{EqDefA2}\end{eqnarray}
 with\begin{equation}
\sigma_{g}^{(1)}\equiv(4\pi)^{5}\alpha_{s}^{3}\alpha^{2}(\sum_{i}e_{i}^{2})^{2}\frac{N_{c}}{N_{c}^{2}-1}.\end{equation}
 The partial amplitudes are summed over all permutations $\sigma$
of the external indices (1,2,3,5) with a fixed cyclic ordering of
(1,2,5), i.e., cyclically-ordered (COP) permutations: \begin{eqnarray}
 &  & \sum_{\sigma\in{\textrm{COP}}_{3}^{(125)}}A_{5;1}(\sigma_{1},\sigma_{2},\sigma_{3},\sigma_{4},\sigma_{5})\equiv\nonumber \\
 &  & \,\,\, A_{5;1}(1,2,5,3,4)+A_{5;1}(1,2,3,5,4)+A_{5;1}(1,3,2,5,4)+A_{5;1}(3,1,2,5,4)+\nonumber \\
 &  & \,\,\, A_{5;1}(5,1,2,3,4)+A_{5;1}(5,1,3,2,4)+A_{5;1}(5,3,1,2,4)+A_{5;1}(3,5,1,2,4)+\nonumber \\
 &  & \,\,\, A_{5;1}(2,5,1,3,4)+A_{5;1}(2,5,3,1,4)+A_{5;1}(2,3,5,1,4)+A_{5;1}(3,2,5,1,4).\label{A51}\end{eqnarray}

The collinear and soft behaviors of the amplitude ${\mathcal{M}}_{5}$
can be established by following the approach in Refs.~\cite{Bern:1993mq,Bern:1993qk,Bern:1994zx,Bern:1994fz,Bern:1998sc,Bern:1999ry,Kosower:1999rx},
extended recently to the two-loop level \cite{Bern:2004cz,Badger:2004uk}.
When gluon 5 is collinear to gluon 1, the amplitude ${\mathcal{M}}_{5}(1,2,3,4,5)$
is dominated by six partial amplitudes with cyclically adjacent indices
5 and 1, such as $A_{5;1}(5,1,2,3,4)$. Similarly, when gluon 5 is
collinear to gluon 2, ${\mathcal{M}}_{5}(1,2,3,4,5)$ is dominated
by six partial amplitudes with cyclically adjacent indices 2 and 5.
Each leading partial amplitude $A_{5;1}(...,5,1,...)$ factors in
the $5\parallel1$ collinear limit into a 4-leg partial amplitude
$A_{4;1}(...,I,...)$ for production of 2, 3, 4, and intermediate
gluon $I$, and amplitude $\mbox{Split}_{-\bar{\lambda}_{I}}^{tree}(5,1)$
\cite{Parke:1986gb,Mangano:1987kp,Berends:1987me,Mangano:1990by}
describing tree-level splitting of $I$ into 5 and 1:\begin{equation}
A_{5;1}(...,5,1,...)\stackrel{5\parallel1}{\longrightarrow}\sum_{\barl_{I}=\pm1}\mbox{Split}_{-\barl_{I}}^{tree}(5,1)A_{4;1}(...,I,...)+\mbox{ subleading terms. }\label{CollinearFactA5}\end{equation}
 The ellipses in Eq.~(\ref{CollinearFactA5}) denote the same permutation
of indices 2, 3, and 4 in $A_{5;1}$ and $A_{4;1}$. The amplitudes
$\mbox{Split}_{-\bar{\lambda}_{I}}^{tree}(5,1)$ are universal functions
of the momenta $\bar{p}_{I},$ $\bar{p}_{1},$ and $\bar{p}_{5}$,
which in our case satisfy $\bar{p}_{I}\equiv\bar{p}_{1}+\bar{p}_{5},$
$\barp_{1}=(1-z)\barp_{I},$ and $\barp_{5}=z\barp_{I},$ where $z=1-1/\widehat{x}_{1}.$
The right-hand side of Eq.~(\ref{CollinearFactA5}) is summed over
the helicities $\bar{\lambda}_{I}$ of $I$. The collinear factorization
relation applies to any one-loop $n-$leg primitive amplitude $A_{n}^{loop}(1,...n)$:\begin{eqnarray}
A_{n}^{loop}(...,a,b,...) & \stackrel{a\parallel b}{\longrightarrow} & \sum_{\barl_{I}}\Biggl[\mbox{Split}_{-\barl_{I}}^{tree}(a,b)A_{n-1}^{loop}(...,I,...)+\mbox{Split}_{-\barl_{I}}^{loop}(a,b)A_{n-1}^{tree}(...,I,...)\Biggr].\label{Splits}\end{eqnarray}
 Equation~(\ref{Splits}) is evaluated here for $n=5$ external legs,
along with the condition that the tree primitive amplitude $A_{4}^{tree}$
vanishes in $0\rightarrow gg\gamma\gamma$ process.

Using Eqs.~(\ref{EqDefA2}), (\ref{A51}), (\ref{CollinearFactA5}),
we derive the approximate form for $|{\mathcal{M}}_{5}|^{2}$ in the
$5\parallel1$ limit: \begin{eqnarray}
 &  & \left|{\mathcal{M}}_{5}(1,2,3,4,5)\right|^{2}\stackrel{5\parallel1}{\longrightarrow}\sigma_{g}^{(1)}\sum_{\barl_{I},\barl_{I}^{\prime}=\pm1}{\mathcal{M}}_{4}^{*}(\barp_{I},\barl_{I}^{\prime};2,3,4)T_{\barl_{I}^{\prime},\barl_{I}}(\widehat{x}_{1}){\mathcal{M}}_{4}(\barp_{I},\barl_{I};2,3,4).\label{A23}\end{eqnarray}
 Here ${\mathcal{\mathcal{M}}}_{4}(I,2,3,4)\equiv\sum_{\sigma\in S_{3}}A_{4;1}(I,\sigma_{2},\sigma_{3},\sigma_{4})$
is the normalized 4-leg amplitude, obtained by summation of the partial
amplitudes $A_{4;1}(I,2,3,4)$ over all possible permutations $S_{3}$
of the legs 2, 3, and 4. The amplitude ${\mathcal{M}}_{4}$ and complex-conjugate
amplitude ${\mathcal{M}}_{4}^{*}$ are evaluated for independent helicities
$\barl_{I}$ and $\bar{\lambda}_{I}^{\prime}$ of $I$. $T_{\barl_{I}^{\prime},\barl_{I}}(\widehat{x}_{1})$
absorbs contributions from the splitting amplitudes:\begin{eqnarray}
T_{\barl_{I},\barl_{I}'}(\widehat{x}_{1}) & \equiv & \sum_{\barl_{1},\barl_{5}=\pm1}\left[\mbox{Split}_{-\barl_{I}}^{tree}(\frac{1}{\widehat{x}_{1}};1,5)\right]^{*}\mbox{Split}_{-\bar{\lambda_{I}}^{\prime}}^{tree}(\frac{1}{\widehat{x}_{1}};1,5).\end{eqnarray}
 In a basis with $\bar{\lambda}=+1$ and $\bar{\lambda}=-1$, $T_{\barl_{I}^{\prime},\barl_{I}}(x)$
is a matrix of the form\begin{eqnarray}
T_{\barl_{I},\barl_{I}'}(x) & = & \frac{2C_{A}}{2xp_{1}\cdot p_{5}}\left(\begin{array}{cc}
\frac{x}{1-x}+\frac{1-x}{x}+x(1-x) & -\frac{1-x}{x}\frac{[51]}{\langle51\rangle}\\
-\frac{1-x}{x}\frac{\langle51\rangle}{[51]} & \frac{x}{1-x}+\frac{1-x}{x}+x(1-x)\end{array}\right).\label{Tlambda}\end{eqnarray}
 The diagonal entries of $T_{\barl_{I},\barl_{I}'}(\widehat{x}_{1})$
give rise to terms proportional to the unpolarized splitting function
$P_{g/g}(\widehat{x}_{1})$ in the asymptotic cross section, with
\begin{equation}
P_{g/g}(\widehat{x}_{1})=2C_{A}\left[\frac{\widehat{x}_{1}}{\left(1-\widehat{x}_{1}\right)_{+}}+\frac{1-\widehat{x}_{1}}{\widehat{x}_{1}}+\widehat{x}_{1}(1-\widehat{x}_{1})\right]+\frac{11N_{c}-2N_{f}}{6}\delta(1-\widehat{x}_{1}),\end{equation}
where $N_{f}$ is the number of active quark flavors. The off-diagonal
entries give rise to terms proportional to the spin-flip splitting
function \begin{equation}
P_{g/g}^{\prime}(\widehat{x}_{1})=2C_{A}(1-\widehat{x}_{1})/\widehat{x}_{1},\end{equation}
 multiplied by the ratio of spinor products $\langle51\rangle\equiv\langle5+|1-\rangle$
and $[51]\equiv\langle5-|1+\rangle$. In a general reference frame,
$\langle51\rangle/[51]$ is a complex phase depending on the azimuthal
separation $\varphi_{1}-\varphi_{5}$ between the gluons 1 and 5.
In the Collins-Soper frame, this phase reduces to $\langle51\rangle/[51]=-1$.%
\footnote{A collinear approximation for $\left|{\mathcal{M}}_{5}\right|^{2}$
is derived in Ref.~\cite{Bern:2002jx} in the framework of the dipole
factorization formalism \cite{Catani:1996vz}. This approximation
agrees with ours up to phases of the off-diagonal terms, which are
not the same as in Eq.~(\ref{Tlambda}). Our expression is shown
upon a closer examination to produce correct phases in an arbitrary
reference frame \cite{DixonPrivate2005}.%
}

Next, we employ explicit expressions for ${\mathcal{M}}_{4}(I,2,3,4)$
from Ref.~\cite{Bern:2001df}, given by products ${\mathcal{M}}_{4}(I,2,3,4)=S_{\barl_{I}\barl_{2}\barl_{3}\barl_{4}}M_{\barl_{I}\barl_{2}\barl_{3}\barl_{4}}^{(1)}$
of reduced matrix elements $M_{\barl_{I}\barl_{2}\barl_{3}\barl_{4}}^{(1)}$
and phase factors $S_{\barl_{I}\barl_{2}\barl_{3}\barl_{4}}$. With
these expressions inserted, Eq.~(\ref{A23}) becomes in the Collins-Soper
frame \begin{equation}
|{\mathcal{M}}_{5}(1,2,3,4,5)|^{2}\stackrel{5\parallel1}{\longrightarrow}\frac{\sigma_{g}^{(1)}}{2\widehat{x}_{1}p_{1}\cdot p_{5}}\Biggl\{ P_{g/g}(\widehat{x}_{1})L_{g}(\theta_{*})+P_{g/g}^{\prime}(\widehat{x}_{1})\, L_{g}^{\prime}(\theta_{*})\,\cos2\varphi_{*}\Biggr\},\label{M5coll51}\end{equation}
 where\begin{equation}
L_{g}(\theta_{*})=\sum_{\bar{\lambda}_{1},\bar{\lambda}_{2},\bar{\lambda}_{3},\bar{\lambda}_{4}}\left|M_{\barl_{I}\barl_{2}\barl_{3}\barl_{4}}^{(1)}\right|^{2},\end{equation}
 and \begin{equation}
L_{g}^{\prime}(\theta_{*})=-4\mbox{{Re}}\left\{ M_{1,1,-1,-1}^{(1)}+M_{1,-1,1,-1}^{(1)}+M_{-1,1,1,-1}^{(1)}+1\right\} .\end{equation}

In the $5\parallel2$ collinear limit $\left|{\mathcal{M}}_{5}\right|^{2}$
is\begin{eqnarray}
 &  & |{\mathcal{M}}_{5}(1,2,3,4,5)|^{2}\stackrel{5\parallel2}{\longrightarrow}\frac{\sigma_{g}^{(1)}}{2\widehat{x}_{2}p_{2}\cdot p_{5}}\Biggl\{ P_{g/g}(\widehat{x}_{2})\, L_{g}(\theta_{*})+P_{g/g}^{\prime}(\widehat{x}_{2})\, L_{g}^{\prime}(\theta_{*})\,\cos2\varphi_{*}\Biggr\}.\label{M5coll52}\end{eqnarray}
 In the soft limit $p_{5}^{\mu}\rightarrow0$, $\left|{\mathcal{M}}_{5}\right|^{2}$
factors as\begin{equation}
\left|{\mathcal{M}}_{5}(1,2,3,4,5)\right|^{2}\rightarrow\frac{1}{Q_{T}^{2}}2C_{A}\frac{\alpha_{s}}{\pi}\left|{\mathcal{M}}_{4}(1,2,3,4)\right|^{2}+\mbox{subleading terms.}\label{M5soft}\end{equation}
 Inserting collinear and soft approximations (\ref{M5coll51}), (\ref{M5coll52}),
and (\ref{M5soft}) in Eq.~(\ref{Asygg1}) and making some simplifications,
we derive the asymptotic expression for real-emission contributions,
\begin{eqnarray}
 &  & \left.\frac{d\sigma_{gg}}{dQ^{2}dy\, dQ_{T}^{2}d\Omega_{*}}\right|_{real}\rightarrow\frac{\sigma_{g}^{(0)}}{S}\frac{1}{2\pi Q_{T}^{2}}\frac{\alpha_{s}}{\pi}\nonumber \\
 & \times & \Biggl\{ L_{g}(\theta_{*})\biggl(f_{g/h_{1}}(x_{1},\mu_{F})f_{g/h_{2}}(x_{2},\mu_{F})\left({\mathcal{A}}_{g}^{(1,c)}\ln\frac{Q^{2}}{Q_{T}^{2}}+{\mathcal{{\mathcal{B}}}}_{g}^{(1,c)}\right)\nonumber \\
 & + & \left[P_{g/g}\otimes f_{g/h_{1}}\right](x_{1},\mu_{F})\, f_{g/h_{2}}(x_{2},\mu_{F})+f_{g/h_{1}}(x_{1},\mu_{F})\left[P_{g/g}\otimes f_{g/h_{2}}\right](x_{2},\mu_{F})\biggr)\nonumber \\
 & + & \cos2\varphi_{*}L_{g}^{\prime}(\theta_{*})\biggl(\left[P_{g/g}^{\prime}\otimes f_{g/h_{1}}\right](x_{1},\mu_{F})\, f_{g/h_{2}}(x_{2},\mu_{F})\nonumber \\
 & + & f_{g/h_{1}}(x_{1},\mu_{F})\left[P_{g/g}^{\prime}\otimes f_{g/h_{2}}\right](x_{2},\mu_{F})\biggr)\Biggr\}.\end{eqnarray}
 Once we add the two-loop 4-leg virtual corrections {[}Fig.~\ref{Fig:FeynDiag}(e)],
the soft singularities in the real-emission cross section residing
at $Q_{T}=0$ are canceled~\cite{Bern:2002jx,Nadolsky:2002gj}. The
final small-$Q_{T}$ expression coincides with Eq.~(\ref{ASYgg}).


\end{document}